\documentclass[runningheads]{llncs}
\usepackage{graphicx}
\PassOptionsToPackage{babel}{english}

\usepackage[utf8]{inputenc}
\usepackage[T1]{fontenc}
\usepackage[american]{babel} 
\usepackage{silence}
\usepackage{fancybox}
\usepackage{adjustbox}

\let\proof\relax\let\endproof\relax 
\usepackage{amsthm}

\usepackage{xcolor}
\usepackage{csquotes} 
\usepackage{siunitx} 
\usepackage{xcolor}
\usepackage{xifthen}
\usepackage[hyphens]{url}
\usepackage{stmaryrd}
\usepackage{mathtools}
\usepackage{nth}
\usepackage{amsbsy}
\usepackage{enumitem}
\usepackage{booktabs}
\usepackage{tabularx}
\usepackage[hidelinks]{hyperref}
\usepackage[capitalize]{cleveref}
\usepackage{subcaption}
\usepackage{algorithm}
\usepackage[noend]{algpseudocode}
\usepackage{float}
\usepackage{datetime2} 
\usepackage{pgfplots}
\usepackage{relsize}
\usepackage{multicol}
\usepackage[nodayofweek]{datetime}
\usepackage{scrhack} 
\usepackage{graphicx}
\usepackage{outlines}
\usepackage{placeins} 
\usepackage{pifont}
\usepackage[normalem]{ulem}

\crefname{section}{\S}{Sections}
\crefname{page}{page}{pages}
\crefname{paragraph}{\S}{Sections}
\Crefname{section}{Section}{Sections}
\crefformat{section}{\S#2#1#3}
\Crefformat{section}{Section~#2#1#3}

\newcommand{\enumfont}{\bfseries}
\SetEnumitemKey{goals}{label={\enumfont P\arabic*}, ref=P\arabic*}
\SetEnumitemKey{goalscore}{label={\enumfont C\arabic*}, ref=C\arabic*}
\SetEnumitemKey{goalsopt}{label={\enumfont Opt}, ref=Opt}
\SetEnumitemKey{goalsadditional}{label={\enumfont S\arabic*}, ref=S\arabic*}
\SetEnumitemKey{goalsS4}{label={\enumfont S4}, ref=S4}
\SetEnumitemKey{freq}{label={\enumfont F\arabic*}, ref=F\arabic*}
\SetEnumitemKey{ereq}{label={\enumfont E\arabic*}, ref=E\arabic*}

\pgfplotsset{compat=1.16}
\graphicspath{{figures/}}

\makeatletter
\newcommand\nobreakpar{\par\nobreak\@afterheading}
\makeatother

\definecolor{color0}{RGB}{0, 0, 0} 
\definecolor{color1}{RGB}{255, 65, 68} 
\definecolor{color1b}{RGB}{252, 91, 99} 
\definecolor{color2}{RGB}{234, 127, 43} 
\definecolor{color2b}{RGB}{253, 174, 97} 
\definecolor{color3}{RGB}{200, 200, 0} 
\definecolor{color3b}{RGB}{255, 255, 191} 
\definecolor{color4}{RGB}{98, 181, 86} 
\definecolor{color4b}{RGB}{171, 221, 164} 
\definecolor{color5}{RGB}{43, 131, 186} 
\definecolor{color5b}{RGB}{158, 204, 239} 
\colorlet{red}{color1}
\colorlet{orange}{color2}
\colorlet{yellow}{color3}
\colorlet{green}{color4}
\colorlet{blue}{color5}

\colorlet{green}{color4}
\colorlet{orange}{color2}
\colorlet{red}{color1}
\definecolor{gray1}{rgb}{.4,.4,.4}

\DeclareSIUnit{\kB}{kB}
\DeclareSIUnit{\KB}{\kB}
\DeclareSIUnit{\MB}{\mega\byte}
\DeclareSIUnit{\mbps}{Mbps}
\DeclareSIUnit{\gbps}{Gbps}
\DeclareSIUnit{\Gbps}{\gbps}
\DeclareSIUnit{\tbps}{Tbps}
\DeclareSIUnit{\kbps}{kbps}

\newcommand{\gma}{GMA}

\newcommand{\pa}{PA}

\newcommand{\pad}{\pa{}$^{\textrm{dist}}$}

\newcommand{\var}[1]{\mathit{#1}}
\newcommand{\Talloc}[4][M]{\ensuremath{{#1}^{\ifthenelse{\isempty{#2}}{}{(#2)}}_{#3\ifthenelse{\isempty{#3}\AND\isempty{#4}}{}{,}#4}}}
\newcommand{\T}[3]{\Talloc{#1}{#2}{#3}}
\newcommand{\Tnew}[3]{\Talloc[\widetilde{M}]{#1}{#2}{#3}}
\newcommand{\tmat}[1]{\T{#1}{}{}}
\newcommand{\ifcap}[2]{\ensuremath{\var{cap}^{\ifthenelse{\isempty{#1}}{}{(#1)}}_{#2}}}
\newcommand{\conv}[2]{\ensuremath{\var{CON}^{\ifthenelse{\isempty{#1}}{}{(#1)}}_{#2}}}
\newcommand{\divr}[2]{\ensuremath{\var{DIV}^{\ifthenelse{\isempty{#1}}{}{(#1)}}_{#2}}}
\newcommand{\convprime}[2]{\ensuremath{\var{\widehat{CON}}^{\ifthenelse{\isempty{#1}}{}{(#1)}}_{#2}}}
\newcommand{\divrprime}[2]{\ensuremath{\var{\widehat{DIV}}^{\ifthenelse{\isempty{#1}}{}{(#1)}}_{#2}}}
\newcommand{\ff}[1]{\ensuremath{\var{f}^{\ifthenelse{\isempty{#1}}{}{(#1)}}}}
\newcommand{\au}[1]{\ensuremath{\var{a}^{(#1)}}}
\newcommand{\bu}[1]{\ensuremath{\var{b}^{(#1)}}}
\newcommand{\xmin}[1][]{\ensuremath{#1{x}^\star}}
\newcommand{\xold}{\xmin}
\newcommand{\xnew}{\xmin[\hat]}
\newcommand{\IF}[1]{\ensuremath{\var{I}^{(#1)}}}
\newcommand{\ffg}[1]{\ensuremath{\var{g}^{\ifthenelse{\isempty{#1}}{}{(#1)}}}}
\newcommand{\pcap}[1]{\ensuremath{\mathcal{A}_{#1}}}
\newcommand{\pcapbrackets}[1]{\ensuremath{\mathcal{A}\ifthenelse{\isempty{#1}}{}{(#1)}}}
\newcommand{\gmaG}[1]{\ensuremath{\mathcal{G}({#1})}}
\newcommand{\gmaGhat}[1]{\ensuremath{\widehat{\mathcal{G}}({#1})}}

\newcommand{\palloc}{path allocation}
\newcommand{\pallocs}{path allocations}

\newcommand{\aone}[1]{\ensuremath{\mathcal{A}_1\ifthenelse{\isempty{#1}}{}{(#1)}}}
\newcommand{\atwo}[2][]{\ensuremath{#1{\mathcal{A}}_2\ifthenelse{\isempty{#2}}{}{(#2)}}}

\Crefformat{section}{Section~#2#1#3}

\renewcommand{\paragraph}[1]{\paragraphnoperiod{#1.}}
\newcommand{\paragraphnoperiod}[1]{\medskip\noindent\textbf{#1}\quad}

\newcommand{\acover}[1]{\ifthenelse{\isempty{#1}}{$\alpha$}{\emph{#1}}-cover}

\newcommand*{\allocmatrix}[1]{allocation matri\ifthenelse{\isempty{#1}}{x}{ces}}
\newcommand*{\Allocmatrix}[1]{Allocation matri\ifthenelse{\isempty{#1}}{x}{ces}}
\newcommand*{\matrixentry}[1]{allocation-matrix entr\ifthenelse{\isempty{#1}}{y}{ies}}
\newcommand*{\pairalloc}[1]{pair allocation\ifthenelse{\isempty{#1}}{}{s}}

\newcommand{\paragraphProof}[1]{%
\paragraphProofNobreak{#1}
}
\newcommand{\paragraphProofNobreak}[1]{\noindent\textbf{#1}}

\newcommand{\subparagraphProofNobreak}[1]{\indent\paragraphProofNobreak{#1}}

\newenvironment{nospaceflalign}[1][4pt]
{\setlength{\abovedisplayskip}{#1}\setlength{\belowdisplayskip}{4pt}%
	\csname flalign\endcsname}
{\csname endflalign\endcsname\ignorespacesafterend}
\newenvironment{nospaceflalign*}
{\setlength{\abovedisplayskip}{5pt}\setlength{\belowdisplayskip}{0pt}%
	\csname flalign*\endcsname}
{\csname endflalign*\endcsname\ignorespacesafterend}

\newtheoremstyle{mytheoremstyle}{5pt}{5pt}{\itshape}{}{\bfseries}{.}{.5em}{}
\theoremstyle{mytheoremstyle}
\newtheorem{thm}{Theorem}
\newtheorem{lem}[thm]{Lemma}
\crefname{thm}{Theorem}{Theorems}
\Crefname{thm}{Theorem}{Theorems}
\crefname{lem}{Lemma}{Lemmas}
\Crefname{lem}{Lemma}{Lemmas}
\newenvironment{myproof}[1][\proofname]{%
  \proof[\noindent\normalfont\bfseries #1]%
}{\endproof}
\newenvironment{subparagraphindent}{%
	\par%
	\leftskip=10pt\rightskip=0pt%
	\noindent\ignorespaces}{%
	\par%
	}
\definecolor{eth1}{HTML}{1F407A}
\definecolor{eth2}{HTML}{3C5A0F}
\definecolor{eth3}{HTML}{0069B4}
\definecolor{eth4}{HTML}{72791C}
\definecolor{eth5}{HTML}{91056A}
\definecolor{eth6}{HTML}{6F6F6E}
\definecolor{eth7}{HTML}{A8322D}
\definecolor{eth8}{HTML}{007A92}
\definecolor{eth9}{HTML}{956013}
\definecolor{eth10}{HTML}{82BE1E}

\usepackage[eulergreek]{sansmath}
\pgfplotsset{
  tick label style = {font=\sansmath\sffamily},
  every axis label = {font=\sansmath\sffamily},
  x label style = {font=\sansmath\sffamily},
  y label style = {font=\sansmath\sffamily},
  legend style = {font=\tiny\sansmath\sffamily},
  label style = {font=\small\sansmath\sffamily}
}

\setlist[itemize]{noitemsep, topsep=3pt plus 3pt minus 3pt}

\DeclareMathOperator*{\argmin}{\text{arg\,min}}

\usepackage[breakable]{tcolorbox}
\tcbsetforeverylayer{autoparskip}
\definecolor{lightgray}{gray}{0.92}
\tcbuselibrary{breakable}

\newcommand{\greybox}[2][]{%
    \begin{tcolorbox}[breakable, colback=lightgray, colframe=lightgray, left=1pt,right=1pt,top=0pt,bottom=0pt]%
        \ifthenelse{\isempty{#1}}{}{\textbf{#1.}\par}%
        #2%
    \end{tcolorbox}%
}

\newenvironment{myexample}[1][Example]%
{%
    \begin{tcolorbox}[breakable, colback=white, colframe=lightgray, left=1pt,right=1pt,top=0pt,bottom=0pt]%
    \itshape
    \ifthenelse{\isempty{#1}}{}{\noindent\textbf{#1}\quad}%
}
{\end{tcolorbox}}

\newcommand{\nF}[1]{\text{\bgroup\sffamily\upshape #1\egroup}}
\newcommand{\nodeExp}[2]{\ensuremath{\nF{#1}^{\nF{\bfseries#2}}}}
\newcommand{\Anode}[1]{\nodeExp{A}{#1}}
\newcommand{\Bnode}[1]{\nodeExp{B}{#1}}
\newcommand{\Cnode}[1]{\nodeExp{C}{#1}}

\theoremstyle{remark}

\theoremstyle{remark}

\newcommand{\appref}[1]{Appendix~#1 of the full paper~\cite{gma-arxiv}} 
\newcommand{\appfigref}[2]{Fig.~#1 in \appref{#2}}
\newcommand{\Appfigref}[2]{Figure~#1 in \appref{#2}}
\newcommand{\iffullversion}[1]{}
\newcommand{\makefullversion}{
    \renewcommand{\iffullversion}[1]{##1}
    \renewcommand{\appref}[1]{\cref{app:##1}}
    \renewcommand{\appfigref}[2]{\cref{fig:##1} in \appref{##2}}
    \renewcommand{\Appfigref}[2]{\Cref{fig:##1} in \appref{##2}}
}

\makefullversion

\begin{document}
\title{\gma{}: A Pareto Optimal Distributed Resource-Allocation Algorithm}
\titlerunning{Global myopic resource allocation}
%
\author{Giacomo Giuliari \and
    Marc Wyss \and
    Markus Legner \and
Adrian Perrig}
%
\authorrunning{G. Giuliari et al.}
%
\institute{ETH Zürich, Universitätstrasse 6, 8092 Zürich, Switzerland\\
\email{\{giacomog, marc.wyss, markus.legner, adrian.perrig\}@inf.ethz.ch}}
\maketitle              
\begin{abstract}


To address the raising demand for strong packet delivery guarantees in
networking, we study a novel way to perform graph resource allocation. We first
introduce \emph{allocation graphs}, in which nodes can independently set
local resource limits based on physical constraints or policy decisions. In this
scenario we formalize the \emph{distributed path-allocation} (\pad{}) problem,
which consists in allocating resources to paths considering only \emph{local}
on-path information---importantly, not knowing which other paths could have an
allocation---while at the same time achieving the \emph{global} property of
never exceeding available resources.


Our core contribution, the global myopic allocation (\gma{}) algorithm, is a
solution to this problem. We prove that \gma{} can compute 
\emph{unconditional} allocations for all paths on a graph, while never
over-allocating resources. Further, we prove that \gma{} is Pareto optimal with
respect to the allocation size, and it has linear complexity in the input size. 
Finally, we show with simulations that this theoretical result could be indeed
applied to practical scenarios, as the resulting path allocations are large
enough to fit the requirements of practically relevant applications.

\end{abstract}

\section{Introduction} \label{sec:intro}

Allocating resources such as bandwidth in a network has proven to be a
difficult problem from both a theoretical and practical perspective: in many
cases, networks consist of independent nodes without central controller and
without a global view of the topology and available resources. Furthermore,
these nodes often have their own policies on how to allocate resources. To the
best of our knowledge, the theoretical networking literature is lacking
solutions that address this distributed setting. In this paper, we consider
\emph{allocation graphs}, directed graphs consisting of independent nodes
augmented with local policies, i.e., the amount of resources each node
allocates for transit between any pair of neighbors. While we interpret the
resources as bandwidth, other interpretations---like computations on behalf of
the neighbors---are possible as well.

For any path in the allocation graph, we want to \emph{myopically} compute a
static allocation, i.e., based only on the local policies of on-path nodes.
This allocation should guarantee that no local allocation is ever exceeded,
even when all path allocations in the network are fully used simultaneously.
This is resource allocation is  therefore \emph{unconditional}, since the size
of one allocation is completely independent of any other allocation, and not
determined by an admission process, and thus cannot be influenced by single
off-path nodes. In particular, nodes do not need to keep track of allocations
as each individual allocation is valid independently of whether or not any
other allocations are used. We formalize the problem of finding the size of
such allocations as the \emph{distributed path allocation} (\pad{}) problem.
Two major questions then arise: (i) Can unconditional resource allocation
indeed be performed in a distributed setting, where nodes have only partial
information on the network, without creating over-allocation? And (ii), since
an allocation is implicitly created for every path in the network, can
allocations be large enough to be useful in practice?

Our work addresses these problems, finding that it is possible to both avoid
over-allocation and create allocations that meet the demands of a number of
modern critical applications at the same time. We show this constructively, by
proposing the first unconditional resource allocation algorithm: the
\emph{global myopic allocation} (\gma{}) algorithm. \gma{} interprets each
node's local allocations both as capacity limits that must not be exceeded and
as policy decisions about the relative importance of links to neighbors. It
efficiently computes allocations that scale with these local policies, and
ensures that capacities are not over-allocated. We prove that \gma{} fulfills
all desired properties and that it is \emph{Pareto optimal} with respect to all
other algorithms that solve the \pad{} problem. Finally, we simulate \gma{} on
random graphs, chosen to model real-world use cases; we evaluate the size of
the resulting path allocations and show that they are viable for practical
applications.

\paragraph{Practical relevance of the \pad{} problem}
Over the past decades, computer networks have predominantly relied on the
\emph{best-effort} paradigm. End-points run congestion-control algorithms to
prevent a congestion-induced collapse of the
network~\cite{Jacobson1988,Kelly1998}, but no further guarantees for packet
delivery or quality of service can be given. This has been shown to work
reasonably well for many applications like web browsing, but it is becoming
increasingly clear that it is far from optimal in terms of performance and
fairness~\cite{Ware2019,Brown2020}.

Although the networking community has developed several protocols to reserve
resources for individual connections~\cite{ATM,RFC1633,Basescu2016SIBRA},
none of them has seen wide-spread adoption because of their high complexity and
poor scalability. These drawbacks arise in all these systems as they offer
\emph{conditional} allocations: endpoints can select the amount of resource to
allocate, the rationale being that supply and demand will eventually lead to
optimal resource utilization. However, this also means that all nodes have to
store information about all individual requests, and check that new requests do
not exceed resource capacity.

An unconditional resource allocation system based on the \gma{} algorithm
avoids this problem. In a network of compliant sources using such a system,
nodes do not need to keep track of allocations as each allocation is valid
independently of whether or not any other allocations are used. Further, \gma{}
guarantees that no over-allocation of bandwidth---and therefore
congestion---occurs. Thus, strong delivery guarantees can be provided to the
communications in this network, without the overhead required by conditional
systems. \appref{A} presents overview of the \emph{critical}
applications that would benefit the most from an unconditional resource
allocation system.

\section{Preliminaries: formalizing resource allocation}
\label{sec:setting-pa-problem}

We now introduce the formalism we use throughout the paper, and characterize
the path-allocation (\pa{}) problem. Although the \pa{} problem arises from an
applied networking context (as some of the terminology also suggests), we seek
to provide a formulation that is not tied to networking, such that our solution
can also be applied to other areas. Therefore, we define the problem with the
abstraction of \emph{allocation} graphs.

\paragraph{Allocation graphs}
We augment the standard directed graph definition, comprising nodes and edges,
with a set of \emph{interfaces} at every node.\footnote{A node can be thought
of as, e.g., an autonomous system in the Internet, or any other entity
part of a distributed system that acts independently from other entities.} An
interface denotes the end of one of the edges attached to a node, while a
\emph{local interface}, which is not associated with any edge, represents
internal sources or sinks (these concepts are shown in
\cref{fig:gma_allocation_graph,fig:gma_alloc_detail} on
\cpageref{fig:gma_allocation_graph}). In an allocation graph, a
\emph{resource}---a generic quantity of interest---is associated with edges,
and is a measure of supply. The \emph{capacity} of an edge is a fixed, positive
real number that represents the maximum amount of resource it can provide;%
\footnote{We use dimensionless values for the resource; in practice, these
could correspond to, e.g., bandwidth (in \si{\gbps}) or computations per
second.}
it is denoted by \ifcap{k}{i, \textrm{IN}}, for the capacity of the edge
incoming to interface $i$ of node $k$, and \ifcap{k}{i, \textrm{OUT}} for
the outgoing edge. Further, we assume that an
\emph{\allocmatrix{}}~\tmat{k} is given for each node~$k$. Allocation
matrices are illustrated in
\cref{fig:gma_alloc_detail,fig:gma_alloc_matrix}. An entry \T{k}{i}{j}
in the \allocmatrix{}, called \textit{\pairalloc{}}, denotes the
maximum amount of resource that can be allocated in total to all the
paths incoming at interface~$i$ and outgoing at interface~$j$.
Allocation matrices are non-negative and not necessarily symmetric. We
call the maximum amount of resource that can be allocated from an
interface $i$ to every other interface the \emph{divergent}, and the
maximum amount of resource that can be allocated from every other
interface towards an interface~$j$ the \emph{convergent}. They are
calculated as the sum of rows or columns of \tmat{k}, respectively:
\begin{nospaceflalign}[5pt]
  \divr{k}{i} & = \sum_{j}\T{k}{i}{j},
    \qquad
  \conv{k}{j}  = \sum_{i}\T{k}{i}{j}.
    \label{eq:div_con}
\end{nospaceflalign}
The matrix \T{k}{}{} must be defined to fulfill $\forall i.\ \divr{k}{i} \leq
\ifcap{k}{i, \textrm{IN}}, \conv{k}{i} \leq \ifcap{k}{i, \textrm{OUT}}$, that
is, neither \divr{k}{i} nor \conv{k}{i} respectively exceed the capacity of
the incoming and outgoing edges, connected to interface~$i$ of node~$k$.

Intuitively, an \emph{interface pair} $(i, j)$ is the logical connection
between two interfaces of a node, and thus a \pairalloc{} expresses the maximum
amount of resource the node is willing to provide from one neighbor to the
next. \Allocmatrix{s} can therefore be seen as a way for nodes to encode
policies on the level of service they want to grant to each pair of neighbors.

In this model, we represent a path of $\ell$ nodes $N^1,\dots,N^\ell$ as a list
of nodes and interface pairs $\pi = [ (N^1, i^1, j^1), (N^2, i^2, j^2), \dots,
(N^{\ell}, i^{\ell}, j^{\ell}) ]$.\footnote{This definition implicitly includes
edges. Also, we assume that the interfaces match, i.e., $j^{(k-1)}$ and
$i^{(k)}$ are interfaces at opposite ends of the same directed edge.} To
simplify the presentation, we will omit the nodes from the list when they are
implicitly clear; we will also use the abbreviation $\T{k}{i}{j}\equiv
\T{N^k}{i^k}{j^k}$. We say that a path is \emph{terminated}, if the first
interface of the first pair and the second interface of the last pair are local
interfaces. Otherwise the path is called \emph{preliminary}. A path is
considered \emph{simple} or \emph{loop-free}, if it contains each node at most once.
Furthermore, we use ~$\pi^k$ to denote the \emph{preliminary prefix-path of
length~$k$} of some terminated path~$\pi$ of length $\ell$ ($\pi^k = [ (i^1,
j^1), (i^2, j^2),\dots, (i^{k}, j^{k}) ]$ for $1 \leq k < \ell$). Finally, we
call a path \emph{valid}, if $\T{1}{i}{j}, \dots, \T{\ell}{i}{j} > 0$,
otherwise it is \emph{invalid}.

\paragraph{The \pa{} problem}
We are interested in the problem of allocating the resource on an allocation
graph to paths. A \emph{\palloc{}} is created when a certain amount of resource
is allocated for that path, exclusively reserving this amount on every edge and
interface pair of the path and thus making it unavailable for any other path.
If the sum of the \pallocs{} traversing an edge exceeds the capacity of the
edge, we say that the edge is \emph{over-allocated}. Similarly, an interface
pair $(i^k, j^k)$ is over-allocated if this sum is larger than its
corresponding \pairalloc{} $\T{k}{i}{j}$.

\greybox{
    Given an allocation graph and information on the \allocmatrix{s}, the \pa{}
    problem is to calculate a \palloc{} for any path $\pi$ in this graph with
    the following constraint:
\begin{enumerate}[goalscore,align=left]
    \item \textbf{No-over-allocation:} For all allocation graphs, even if there
        is an allocation on every possible valid path in the graph, no edge and
        no interface pair is ever over-allocated.\footnotemark
    \label{req-no-overuse}
\end{enumerate}
}

\footnotetext{\label{fn:overallocation}Paths with loops, and of arbitrary
length, are also included in this definition.}

\noindent Solving the \pa{} problem then requires finding an algorithm \pcapbrackets{}
that can compute such \pallocs{}. We intentionally left underspecified the
precise input that such an algorithm receives, as it depends on whether the
algorithm is centralized or distributed. If centralized, \pcapbrackets{}'s
input is the whole network topology, as well as the \allocmatrix{s} of all the
nodes. Thus, the centralized \pa{} problem can be viewed as a variant of the
multicommodity flow problem~\cite{ford1958suggested}, with the additional
constraint that \pairalloc{s} have to be respected.

In the distributed version of the \pa{} problem (\pad{}), the algorithm has to
run consistently on each node, with partial information about the allocation
graph. Since nodes on a path are assumed to be able to exchange information, we
restrict this information by requiring \pcapbrackets{}'s input to contain only
information about the path for which the \palloc{} is computed. This is
captured by the following definition:

\greybox{
    The \pad{} problem is to solve the \pa{} problem with this additional restriction:
\begin{enumerate}[goalscore,align=left]
  \setcounter{enumi}{1}
  \item \textbf{Locality:}
  The path allocation is a function of the on-path \allocmatrix{s}~$\T{1}{}{}, \dots, \T{\ell}{}{}$ only.
    \label{req-locality}
\end{enumerate}
}

\noindent Among the set of algorithms that fit this definition, we are naturally
interested in the ones that lead to higher \pallocs{}. Since a precise
optimality condition on the algorithm depends on the practical application for
which it is used, we generally postulate that meaningful algorithms provide
\pallocs{} that cannot be strictly increased. This is captured by
Pareto optimality:
\greybox{
\begin{enumerate}[goalsopt, align=left]
  \item \textbf{Optimality:}
      Consider the class of all algorithms fulfilling the requirement of either \pa{} or \pad{}.
      Algorithm \pcapbrackets{} from this class is (Pareto) optimal if there is no other algorithm $\mathcal{B}$ from the same class that can provide at least the same \palloc{} for every path of every allocation graph, and a strictly better allocation for at least one path.
      Formally, if there exists a graph with a path $\pi$ for which $\mathcal{B}(\pi) = \mathcal{A}(\pi) + \delta$ with $\delta > 0$, then there exists at least one other path $\pi'$, possibly in a different graph, where $\mathcal{B}(\pi') = \mathcal{A}(\pi') - \delta'$ with $\delta' > 0$.\footnotemark
    \label{req-optimality}
\end{enumerate}
}
\footnotetext{The loose constraint that $\pi'$ is possibly in a different graph comes from the fact that because of the locality property in the \pad{} problem, the algorithm has no way to differentiate two graphs having a path with the same nodes and \allocmatrix{s}.}

\noindent In addition, we specify three supplementary properties that make an algorithm
more amenable to practical settings. First, the algorithm should provide
non-zero allocations for all valid paths, second, we require the algorithm to
be efficient in the length of the path and the size of the on-path allocation
matrices, and lastly, we enforce stricter requirements on the policy of
individual nodes with the \emph{monotonicity} property: if a node increases one
of its \pairalloc{s}, we expect all \pallocs{} crossing the interface pair to
at least not decrease.
Increasing one \pairalloc{} also increases the corresponding divergent and
convergent, while all other \pairalloc{s} that contribute to this convergent or
divergent remain the same. Therefore the relative contribution of the increased
\pairalloc{} becomes higher, while the relative contribution of the other
\pairalloc{s} decreases. This way, a node's \allocmatrix{} can also be
understood as a policy that defines the relative importance of its neighbors.
Since a path containing loops might traverse the same node both through a
\pairalloc{} with increased importance and through one with decreased importance,
monotonicity is only meaningful in the context of simple paths. \footnote{For
$i^k_1 \neq i^k_2$, increasing \T{k}{i_1}{j} decreases the relative contribution of
\T{k}{i_2}{j} (\cref{eq:div_con}).}

\greybox{
\begin{enumerate}[goalsadditional,align=left]
  \item \textbf{Usability:} For every valid path~$\pi$, the resulting allocation is positive ($\pcapbrackets{\pi} > 0$).
    \label{req-usability}

\item \textbf{Efficiency:} Algorithm $\mathcal{A}$ should have at most polynomial complexity as a function of input size. Specifically, for \pad{} this means polynomial in the total  size of the allocation matrices of on-path nodes. This is a relatively loose requirement, we will show a linear algorithm in the following.
    \label{req-efficiency}

  \item \textbf{Monotonicity:} If the \pairalloc{} of some node $k$ on a simple path $\pi$ is increased and all other allocations remain unchanged, the resulting allocation must not decrease:
  $\allowbreak\T{k}{i}{j} \leq \Tnew{k}{i}{j} \implies \pcapbrackets{\pi} \leq \widetilde{\mathcal{A}}(\pi)$.
    \label{req-monotonicity}
\end{enumerate}
}

\noindent The challenge of devising an optimal \pad{} algorithm is clear: \pcapbrackets{}
can only rely on a \emph{myopic} view of the path, without any further
knowledge about the larger graph. However, it has to achieve the \emph{global}
constraints of Pareto-optimality and no-over-allocation, which consider the
result of performing allocations on all valid paths.
In the remainder of the paper, we present the global myopic allocation (\gma{})
algorithm as a solution to the \pad{} problem. \gma{} fulfills requirements
\labelcref{req-no-overuse,req-locality}, and is optimal according to
\labelcref{req-optimality}, which we formally prove in \cref{sec:properties}.
Furthermore, we prove in \appref{E} that \gma{} also
satisfies all the supplementary requirements
(\labelcref{req-efficiency,req-monotonicity,req-usability}).
An additional property, \emph{extensibility}, is presented and proven in \appref{F}.

\begin{figure}[thb!]
    \centering
    \begin{subfigure}[t]{0.3\linewidth}
        \centering
        \includegraphics[height=6cm]{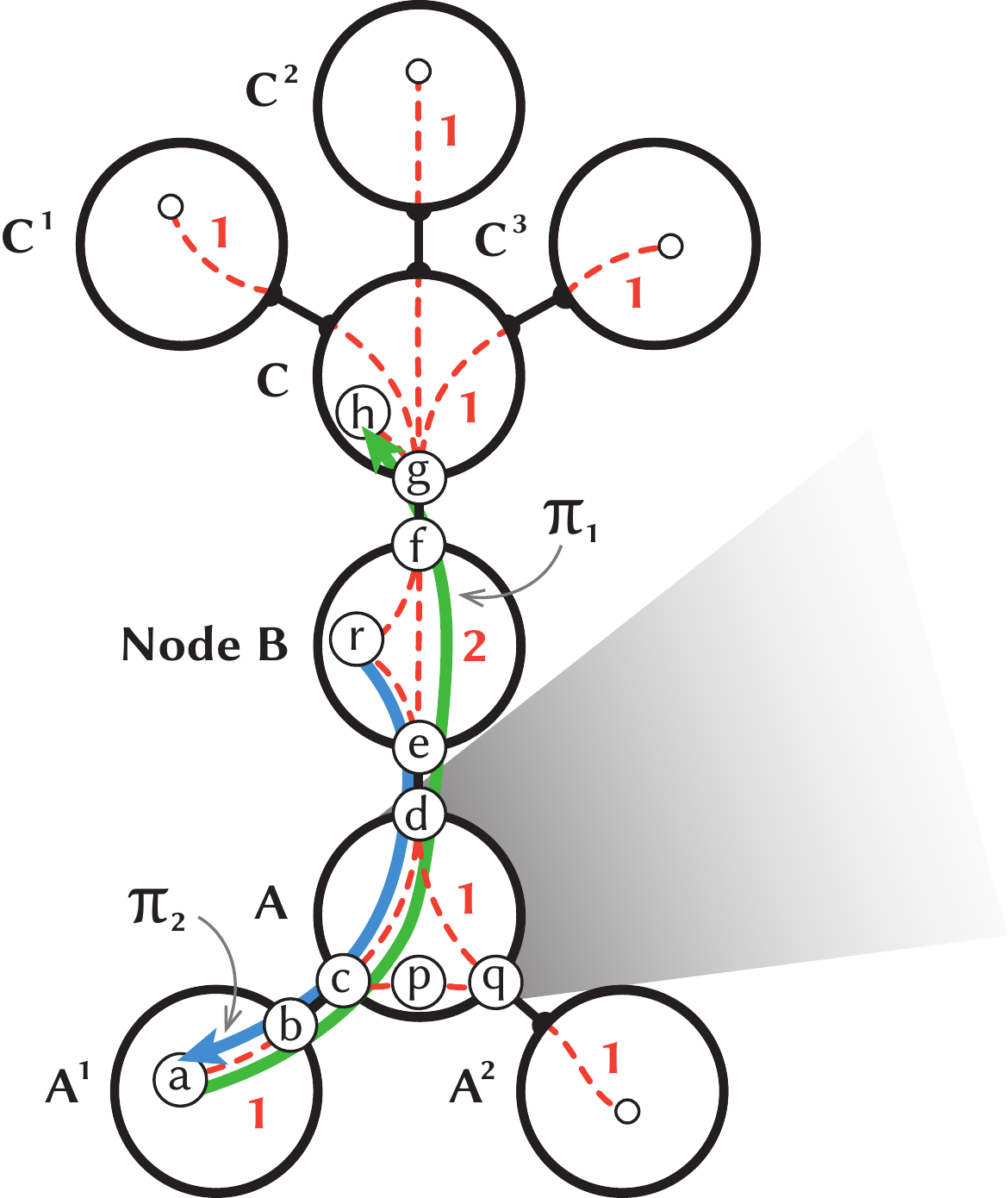}
        \caption{An allocation graph.}
        \label{fig:gma_allocation_graph}
    \end{subfigure}
    \begin{subfigure}[t]{0.29\linewidth}
        \centering
        \includegraphics[height=5cm]{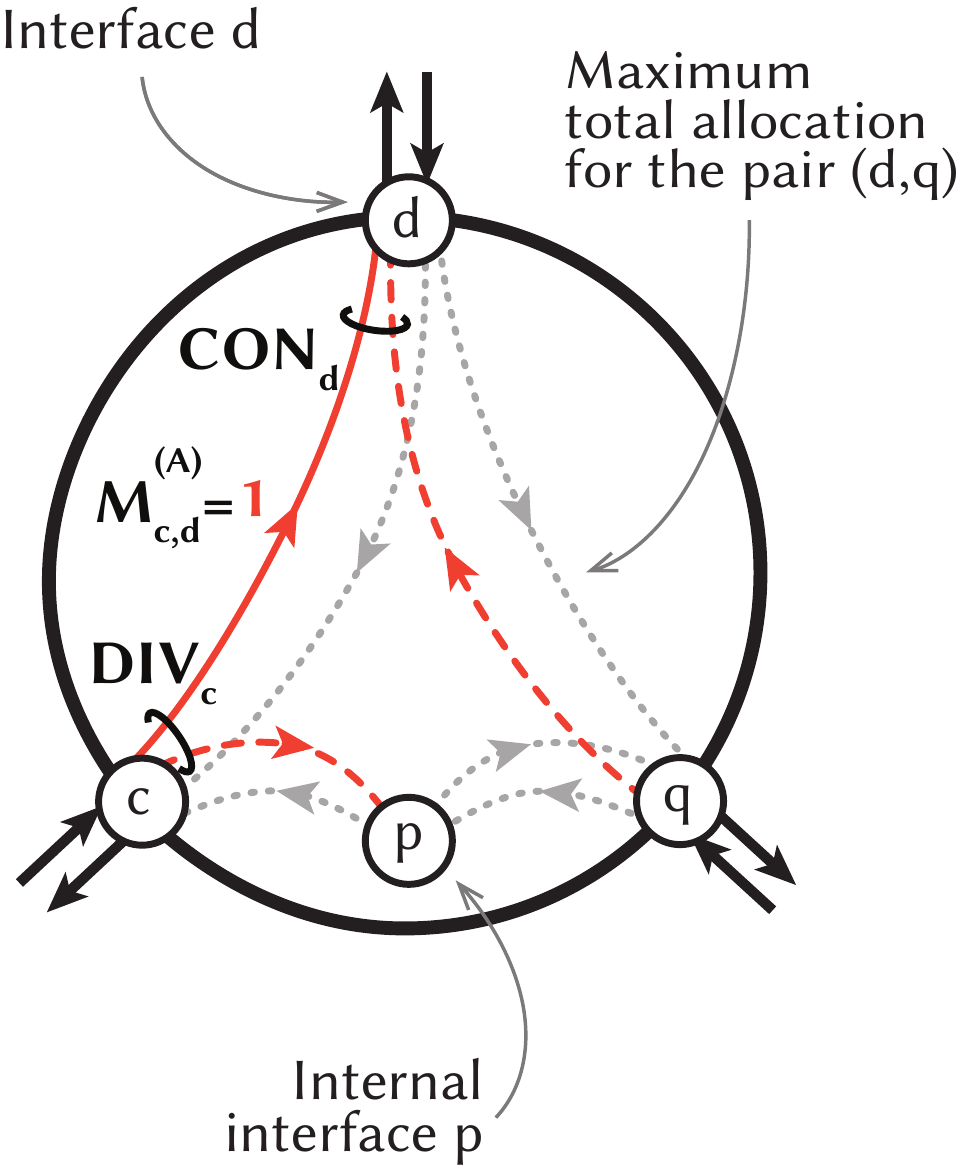}
        \caption{Detailed structure of~\Anode{}.}
        \label{fig:gma_alloc_detail}
    \end{subfigure}
\begin{subfigure}[t]{0.29\linewidth}
        \centering
        \includegraphics[height=4cm]{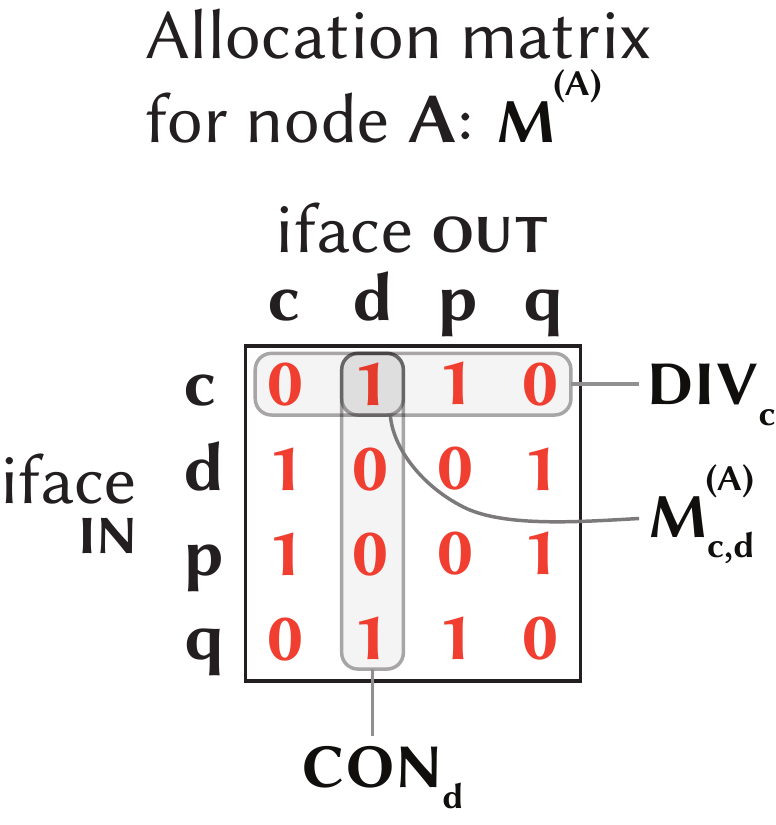}
        \caption{Allocation matrix of~\Anode{}.}
        \label{fig:gma_alloc_matrix}
    \end{subfigure}
    \caption{\textbf{Example of an allocation graph}.
        Pair allocations are represented in \cref{fig:gma_allocation_graph} by
        dashed lines---their size shown by the number in the respective node.
        If two interfaces are not connected by dashed lines, their
        \protect\pairalloc{} is zero. All \protect\pairalloc{s} are
        bidirectional, as shown in \cref{fig:gma_alloc_detail}. For clarity, we
        use globally unique interface identifiers.
        \Cref{fig:gma_allocation_graph} also shows paths $\pi_1$ and $\pi_2$,
        used in the examples ($\pi_3$ is the reverse of $\pi_2$).
  }
    \label{fig:gma_intuition}
\end{figure}

\section{Introducing the GMA algorithm}\label{sec:intuition_solution}

We present the \gma{} algorithm in three steps: starting from a simple first-cut approach, at each step we present a refinement of the previous algorithm.
This section is meant to provide a profound yet intuitive understanding of the \gma{} algorithm and its properties---accompanied by the example in \cref{fig:gma_allocation_graph}---leading to the final formulation of \gma{} in \cref{eq:gma}.

\subsection{Step 1: towards no-over-allocation}
As a first attempt to achieve no-over-allocation, we take the \pairalloc{} of the first node on a path, and multiply it by the ratio of the \pairalloc{} and the divergent for each of the traversed interface pairs.
With this approach, each node receives a \emph{preliminary allocation} from the previous node, fairly splits it among all interfaces according to the pair allocations, and passes it on to the next node.
This leads to the following formula:
\begin{nospaceflalign}
  \aone{\pi} = \T{1}{i}{j} \cdot \prod_{k = 2}^{\ell} \frac{\T{k}{i}{j}}{\divr{k}{i}}.
    \label{eq:fair_allocation_split}
\end{nospaceflalign}
\begin{myexample}
    Consider the path $\pi_1=[(\Anode{1}, \nF a, \nF b), (\Anode{}, \nF c, \nF d), (\Bnode{}, \nF e, \nF f), (\Cnode{}, \nF g, \nF h)]$ in \cref{fig:gma_allocation_graph}. Then, \cref{eq:fair_allocation_split} results in an allocation $\aone{\pi_1}= 1\cdot\tfrac12\cdot\tfrac24\cdot\tfrac14 = \tfrac{1}{16}$.
\end{myexample}
\noindent To understand the idea behind this formula we consider some node~$k$ with interface $i$, connected through this interface to a neighboring node~$n$. If node~$n$ can guarantee that the sum the preliminary allocations of all preliminary paths going towards node~$k$ is at most \divr{k}{i}, then \aone{} ensures that for each of node~$k$'s interfaces~$j$, the sum of all preliminary allocations of all preliminary paths going through $(i, j)$ is at most $\T{k}{i}{j}$. If all neighbors can provide this guarantee, no \pairalloc{} of node~$k$ will be over-allocated, which implies that also none of its convergents will be over-allocated. If node~$k$'s convergents are smaller or equal to the corresponding divergents of its neighbors, also node~$k$ can give this guarantee to all of its neighbors.
Therefore \aone{} will never cause over-allocation, if \emph{every node's convergents are smaller or equal to the corresponding divergents of its neighbors}---which is an assumption we want to get rid of.

\begin{myexample}
    The graph in \cref{fig:gma_allocation_graph} ensures that the divergent of a node is always larger than the convergent of the previous node when going \emph{upwards}. Going \emph{downwards}, this is not the case. Indeed, already two paths $\pi_2=[(\Bnode{}, \nF r, \nF e), (\Anode{}, \nF d, c), (\Anode{1}, \nF b, \nF a)]$ with $\aone{\pi_2} = 2\cdot\tfrac12\cdot\tfrac11=1$ and $\pi_3=[(\Cnode{}, \nF h, \nF g), (\Bnode{}, \nF f, \nF e), (\Anode{}, \nF d, c), (\Anode{1}, \nF b, \nF a)]$ (reverse of $\pi_1$) with $\aone{\pi_3} = 1\cdot\tfrac24\cdot\tfrac12\cdot\tfrac11=\tfrac14$ together cause an over-allocation of interface pairs $(\nF d, \nF c)$ and $(\nF b, \nF a)$.
\end{myexample}

\subsection{Step 2: a general solution for no-over-allocation}
As over-allocation with \aone{} can only occur when some node's convergent is larger than the corresponding divergent of its neighbor, we can normalize each preliminary allocation to compensate this disparity.
More concretely, if $\conv{k-1}{i} > \divr{k}{j}$ for an on-path node $k$, the preliminary allocation from node~$k-1$ is multiplied with:
\begin{nospaceflalign}
  \frac{\divr{k}{j}}{\conv{k-1}{i}} \cdot \frac{\T{k}{i}{j}}{\divr{k}{j}} = \frac{\T{k}{i}{j}}{\conv{k-1}{i}}.
    \label{eq:compensate}
\end{nospaceflalign}
Adapting \cref{eq:fair_allocation_split} to this modification gives rise to the following formula:
\begin{nospaceflalign}
  \atwo{\pi} = \T{1}{i}{j} \cdot \prod_{k = 2}^{\ell} \frac{\T{k}{i}{j}}{\max\{\conv{k-1}{j}, \divr{k}{i} \}}.
    \label{eq:gma_old}
\end{nospaceflalign}
\begin{myexample}
    We find $\atwo{\pi_3}= 1\cdot\tfrac24\cdot\tfrac14\cdot\tfrac12 = \tfrac{1}{16}=\atwo{\pi_1}$; $\atwo{\pi_2}=2\cdot\tfrac14\cdot\tfrac12 = \tfrac{1}{4}$.
\end{myexample}
\noindent This algorithm will never cause over-allocation, which follows directly from our proof in \cref{sec:no-overuse}.
Unfortunately, \atwo{} is neither monotonic nor Pareto optimal.
We can see why this is the case by taking a closer look at the contribution of some node~$k$ to the calculated allocations, which consists of the values $(\divr{k}{i}, \T{k}{i}{j},\allowbreak \conv{k}{j})$.
In \cref{eq:gma_old}, the only subterm depending on those values is
\begin{nospaceflalign}
  \frac{\T{k}{i}{j}}{\max\{\conv{k-1}{j}, \divr{k}{i} \} \cdot \max\{\conv{k}{j}, \divr{k+1}{i} \}}.
    \label{eq:contribution}
\end{nospaceflalign}
Increasing $\T{k}{i}{j}$ by $\delta > 0$, and thus, implicitly, also $\divr{k}{i}$ and $\conv{k}{j}$ by~$\delta$, can potentially contribute twice to the denominator and only once to the nominator of \cref{eq:contribution}, thereby reducing all the allocations going through the interface $(i, j)$.
\begin{myexample}
    Consider increasing the pair allocation $(\nF c, \nF d)$ to $\Tnew{\Anode{}}{\nF c}{\nF d}=9$, leaving everything else unchanged. Then, $\atwo[\widetilde]{\pi_2}=2\cdot\tfrac{9}{10}\cdot\tfrac{1}{10} = \tfrac{18}{100} < \tfrac{1}{4} = \atwo{\pi_2}$.
\end{myexample}
\noindent In general, \atwo{} provides suboptimal allocations when there is a node $k$ with ``\emph{superfluous} allocations'', i.e., where $\divr{k}{i} > \conv{k-1}{j}$ and $\conv{k}{j} > \divr{k+1}{i}$. We explain how to strictly improve this and present \gma{} in the next section.

\subsection{Step 3: monotonic and Pareto-optimal allocations}\label{sec:gma.final}

The main idea to resolve the violation of monotonicity and optimality is to implicitly \emph{scale down} the three-tuple of a node $k$ with superfluous allocations to $(s \cdot\divr{k}{i}, s \cdot\T{k}{i}{j}, s \cdot\conv{k}{j})$ for $0 < s < 1$, such that either $s \cdot\divr{k}{i} \leq \conv{k-1}{j}$ or $s \cdot\conv{k}{j} \leq \divr{k+1}{i}$.
The intuition is that a third algorithm, based on \atwo{} but with scaled-down three-tuples, does not cause over-allocation while observing monotonicity. We will prove later in \cref{sec:properties} that this statement holds.

For some arbitrary path, we now want to find a way to optimally scale down the three-tuple $(\divr{k}{i}, \T{k}{i}{j}, \conv{k}{j})$ of each node $k$. The result is a new algorithm that takes the original inputs, scales them down implicitly, and finally uses \atwo{} to compute the allocation.

As we prove in \appref{B}, down-scaling improves the resulting \palloc{} only for the case---as considered above---in which
superfluous allocations are present ($\divr{k}{i} > \conv{k-1}{j}$ and $\conv{k}{j} > \divr{k+1}{i}$).\footnote{\conv{k-1}{j} and \divr{k+1}{i} might have already been scaled down.}
It is therefore sufficient to scale down the divergent of node $k$ to the convergent of node $k-1$, any further scaling will not improve the allocation.
This observation results in the following iterative algorithm.

On a path $\pi$ with $\ell$ nodes, we start from node $1$. As there is no previous node, scaling is not possible, and the scaling factor is $\ff{1} = 1$.
At the second node, the convergent of the first node can either be smaller than the divergent of the second node, or larger.
In the first case, we scale down the three-tuple of the second node by $\conv{1}{j} / \divr{2}{i}$.
In the second, no scaling down is possible. In both cases we thus scale down the three-tuples of node~$2$ by $\ff{2} = \min\{1, \conv{1}{j} / \divr{2}{i}\}$, and so the first factor of the product in \cref{eq:gma_old} becomes
\begin{nospaceflalign}
  \frac{\T{2}{i}{j} \cdot \ff{2}}{\max\{\conv{1}{j}, \divr{2}{i} \cdot \ff{2} \}} = \frac{\T{2}{i}{j} \cdot \ff{2}}{\conv{1}{j}}.
\end{nospaceflalign}
At the third node this case distinction is repeated. However, recall that the convergent of the second node might have been scaled down, so we have to use the value $(\ff{2} \cdot\conv{2}{j})$ instead of $\conv{2}{j}$ in the computation. Therefore, taking $\ff{3} = \min\{1,~(\conv{2}{j} \cdot \ff{2}) / \divr{3}{i}\}$, we obtain the third factor of the product in \cref{eq:gma_old}:
\begin{nospaceflalign}
  \frac{\T{3}{i}{j} \cdot \ff{3}}{\max\{\conv{2}{j} \cdot \ff{2}, \divr{3}{i} \cdot \ff{3} \}} = \frac{\T{3}{i}{j} \cdot \ff{3}}{\conv{2}{j} \cdot \ff{2}}.
\end{nospaceflalign}
Continuing this expansion, we can define the scaling factors~$f$ recursively for each node as
\begin{nospaceflalign}
  \ff{1} = 1; \qquad  \ff{k} = \min \biggl\{1,~ \frac{\conv{k-1}{j}\cdot \ff{k-1}}{\divr{k}{i}}  \biggr\}.
    \label{eq:gma-f-recursive}
\end{nospaceflalign}
Overall, we modify \cref{eq:gma_old} in the following way:
\begin{nospaceflalign}
  \gmaG{\pi}
  &= \T{1}{i}{j} \cdot \prod_{k=2}^{\ell} \frac{ \T{k}{i}{j} \cdot \ff{k}}{\conv{k-1}{j} \cdot \ff{k-1}}
  = \ff{\ell}\cdot\frac{\prod_{k=1}^{\ell} \T{k}{i}{j}}{\prod_{k=2}^{\ell} \conv{k-1}{j}},
    \label{eq:gma-f}
\end{nospaceflalign}
which is equivalent to computing \atwo{} on the scaled-down input three-tuples.
The last step follows from rearranging indices and realizing that \ff{k} can be factored out recursively, apart from the first ($\ff{1} = 1$) and the last one.
Instead of this recursive formulation, \cref{eq:gma-f} can also be written as a direct formula (the proof can be found in \appref{C}).

\greybox{
  The \emph{global myopic allocation (GMA) algorithm}:
    \begin{nospaceflalign}
      \gmaG{\pi}
      &= \min_{x} \left(\prod_{k=1}^{x-1} \frac{\T{k}{i}{j}}{\conv{k}{j}}\cdot \T{x}{i}{j} \cdot \prod_{k=x+1}^{\ell} \frac{\T{k}{i}{j}}{\divr{k}{i}} \right)
        \label{eq:gma}
    \end{nospaceflalign}
}

\begin{myexample}
    Consider again our example of \cref{fig:gma_allocation_graph} with $\Tnew{\Anode{}}{\nF c}{\nF d}=9$. In this case we have $\divr{\Anode{}}{\nF d} = 10 > \conv{\Bnode{}}{\nF e} = 4$ and $\conv{\Anode{}}{\nF c} = 10 > \divr{\Anode{1}}{\nF b} = 1$. The three-tuple of \Anode{} can thus be scaled down by a factor of $\tfrac{4}{10}$. Using \cref{eq:gma} for the path $\pi_2$, we find that the argument of the minimum is \Anode{1} and $\gmaG{\pi_2}=\tfrac24\cdot \frac{9}{10} \cdot 1 = \frac{9}{20} > \frac{18}{100} = \atwo[\widetilde]{\pi_2}$.
\end{myexample}

\section{Proofs of GMA's properties} \label{sec:properties} \allowdisplaybreaks

In this section, we prove that \gma{}'s computation described in~\cref{eq:gma} satisfies the properties defined in~\cref{sec:setting-pa-problem}.
We prove the core property~\labelcref{req-no-overuse} in \cref{sec:no-overuse} and~\labelcref{req-optimality} in~\cref{sec:optimality}.
Locality (\labelcref{req-locality}) follows directly from \cref{eq:gma}, as the computation only involves \matrixentry{s} of the nodes on the path.
The supplementary properties~\labelcref{req-usability,req-efficiency,req-monotonicity} are proven in~\appref{E}.

\subsection{Proof of no-over-allocation (\labelcref{req-no-overuse})} \label{sec:no-overuse}
In this subsection we prove that there is no resource overuse of any of the \pairalloc{} \T{k}{i}{j}, which, by the fact that convergent and divergent of an interface must be smaller than the capacity of the edge connected to it, implies that there is also no overuse on any edge of the graph.
In the context of this proof, the $+$ operator is not only used for addition, but also for list concatenation.
We denote the set of non-local interfaces of some node $k$ as $\IF{k}_{\text{ext}}$.
We will use the notation $\T{k}{i}{j}(\pi)$ to state more precisely which path the variable refers to.
We want to prove that for every node $k$ and all of its interface pairs, the corresponding \pairalloc{} is greater than or equal to the sum of all resource allocations of all paths going through that interface pair.
For this we distinguish the following cases an interface pair can be assigned to, and prove each case individually:
\begin{enumerate}[noitemsep,topsep=0pt,leftmargin=*,align=left]
    \item[\emph{Case 1:}] The interface pair starts from a local interface: $(\bot, j)$
    \item[\emph{Case 2:}] The interface pair ends in a local interface: $(i, \bot)$
    \item[\emph{Case 3:}] The interface pair starts and ends in non-local interfaces: $(i, j)$
\end{enumerate}
\smallskip

\paragraphProofNobreak{Case 1:}
We will prove a stronger statement, captured by the following lemma:

\begin{lem} \label{lemma:no-overuse-proof}%
  For an arbitrary node $A$ and an arbitrary non-local interface $j^A$, let $S_\mathrm{t}^x$ be the set of \emph{terminated} paths of length \emph{at most} $x$ that start in $(\bot, j^A)$, and $S_\mathrm{p}^x$ the set of \emph{preliminary} paths of length \emph{exactly} $x$ that start in $(\bot, j^A)$. Then
  \begin{nospaceflalign}[5pt]
    \forall x\geq 1: \sum_{\pi \in S_\mathrm{p}^x} \gmaG{\pi} + \sum_{\pi \in S_\mathrm{t}^x} \gmaG{\pi} \leq \T{A}{\bot}{j}.
  \end{nospaceflalign}%
\end{lem}%
\noindent We emphasize that, by the definition in \cref{eq:gma}, \gma{} not only allows to calculate allocations on terminated, but also on preliminary paths.
The lemma implies our original statement, i.e., $\forall x\geq 1$: $\sum_{\pi \in S_\mathrm{t}^x} \gmaG{\pi} \leq \T{A}{\bot}{j}$.

\begin{myproof}
  We prove \cref{lemma:no-overuse-proof} by induction over $x$ for arbitrary $A$ and $j^A$.

  \paragraphProof{\\Base case ($x=1$):}
  \noindent We have $S_\mathrm{p}^1 = \{~[ (\bot, j^A) ] ~\}$ and $S_\mathrm{t}^1 = \{\}$, which directly implies $\sum_{\pi \in S_\mathrm{p}^1} \gmaG{\pi} + \sum_{\pi \in S_\mathrm{t}^1} \gmaG{\pi} = \T{A}{\bot}{j} \leq \T{A}{\bot}{j}$. \medskip

  \paragraphProofNobreak{Inductive step:}
  \begin{subparagraphindent}
    \paragraphProof{Induction hypothesis:}
    For a particular $x$: $\sum_{\pi \in S_\mathrm{p}^x} \gmaG{\pi} + \sum_{\pi \in S_\mathrm{t}^x} \gmaG{\pi} \leq \T{A}{\bot}{j}$. \smallskip \\
    \paragraphProof{To show:}
    $\sum_{\pi \in S_\mathrm{p}^{x+1}} \gmaG{\pi} + \sum_{\pi \in S_\mathrm{t}^{x+1}} \gmaG{\pi} \leq \T{A}{\bot}{j}$.\smallskip \\
    \paragraphProof{Definitions:}
    For some preliminary path $\pi$ of length $\ell$, let node $Z$ be the node that is connected to $j^\ell$ and let the corresponding interface of $Z$ be $i^Z$.
    We define the \emph{local extension of a path $\pi$} as
    $E_{\text{loc}}(\pi) := \{~\pi + [(i^Z, \bot)]~\}$,
    the \emph{non-local extension of a path $\pi$} as
    $E_{\text{ext}}(\pi) := \cup_{j^Z \in \IF{Z}_{\text{ext}}} \{~\pi + [(i^Z, j^Z)]~\}$ and their union as $E(\pi) := E_{\text{loc}}(\pi) \cup E_{\text{ext}}(\pi)$.
  \end{subparagraphindent} \smallskip

	\subparagraphProofNobreak{Proof:}
	\allowdisplaybreaks
  \begin{subequations}
  	\begin{nospaceflalign}[0pt]
    	&\smashoperator{\sum_{\pi \in S_\mathrm{p}^{x+1}}} \gmaG{\pi} + \smashoperator{\sum_{\pi \in S_\mathrm{t}^{x+1}}} \gmaG{\pi}~
    	= \left(\sum_{\pi \in S_\mathrm{p}^{x}} \smashoperator[r]{\sum_{{\phi \in E_{\text{ext}}(\pi)}}} \gmaG{\phi}\right) + \left(\smashoperator[r]{\sum_{\pi \in S_\mathrm{t}^{x}}} \gmaG{\pi} + \smashoperator[l]{\sum_{\pi \in S_\mathrm{p}^{x}}} \smashoperator[r]{\sum_{\phi \in E_{\text{loc}}(\pi)}} \gmaG{\phi}\right) \\
    	&= \smashoperator[l]{\sum_{\pi \in S_\mathrm{p}^{x}}} \smashoperator[r]{\sum_{\phi \in E(\pi)}} \gmaG{\phi} + \smashoperator{\sum_{\pi \in S_\mathrm{t}^{x}}} \gmaG{\pi}
      \label{eq:no-overuse-path-extension}\\
      &= \smashoperator[l]{\sum_{\pi \in S_\mathrm{p}^{x}}} \smashoperator[r]{\sum_{\phi \in E(\pi)}} \min \left(\gmaG{\pi}\cdot \frac{\T{Z}{i}{j}}{\divr{Z}{i}},~\prod_{k=1}^{\ell} \frac{1}{\conv{k}{j}}\cdot\T{Z}{i}{j}\right) + \smashoperator[r]{\sum_{\pi \in S_\mathrm{t}^{x}}} \gmaG{\pi}
      \label{eq:no-overuse-extended-path}\\
      &\leq \smashoperator[l]{\sum_{\pi \in S_\mathrm{p}^{x}}} \smashoperator[r]{\sum_{\phi \in E(\pi)}} ~~\frac{\T{Z}{i}{j}}{\divr{Z}{i}} \cdot \gmaG{\pi} + \smashoperator{\sum_{\pi \in S_\mathrm{t}^{x}}} \gmaG{\pi}
      \label{eq:no-overuse-leq}
    	= \smashoperator{\sum_{\pi \in S_\mathrm{p}^{x}}} \gmaG{\pi} + \smashoperator[r]{\sum_{\pi \in S_\mathrm{t}^{x}}} \gmaG{\pi} ~\leq~ \T{A}{\bot}{j}
    \end{nospaceflalign}
  \end{subequations}
	\begin{subparagraphindent}
	In the step from \cref{eq:no-overuse-path-extension} to \cref{eq:no-overuse-extended-path}, we used the fact that when extending the path, the argument of the minimum of \cref{eq:gma} either stays the same, or the newly added node now minimizes the formula, which follows directly from \cref{eq:gma-f}. The transition in \cref{eq:no-overuse-leq} follows from $\sum_{\phi \in E(\pi)} \T{Z}{i}{j}=\divr{Z}{i}$.
  \end{subparagraphindent}
\medskip

\paragraphProofNobreak{Case 2:}
The proof is exactly the same as for case 1, except that we extend the path in the backward instead of the forward direction. The only change required is the adaptation of the definitions of local and non-local extensions of a path and we use $\sum_{\phi \in E(\pi)} \T{Z}{i}{j}=\conv{Z}{j}$.%
\medskip

\paragraphProofNobreak{Case 3:}
Choose an arbitrary node $A$. Then choose arbitrary non-local interfaces $i^A, j^A \in \IF{A}_{\text{ext}}$ of node $A$.
Using exactly the same procedure as for the proof of case 2, but using $(i^A, j^A)$ as the interface pair where the paths \enquote{end} (it does not terminate in a local interface), we can show that the sum of all resource allocations for all paths \emph{ending} in $(i^A, j^A)$ is always smaller or equal to $\T{A}{i}{j}$.
We then choose an arbitrary path $\pi$ that ends in $(i^A, j^A)$.
Using the same procedure as for the proof of case 1, but using $(i^A, j^A)$ as the interface pair where the paths \enquote{begin} (it does not start in a local interface) and setting $\hat{M}^{(A)}_{i, j} := \gmaG{\pi}$, we can show that the sum of the resource allocations of all the (terminated) paths that extend $\pi$ never exceeds $\gmaG{\pi}$.
It follows that the sum of the resource allocations of all the paths going through $(i^A, j^A)$ never exceeds $\T{A}{i}{j}$.
\end{myproof}


\subsection{Proof of optimality (\labelcref{req-optimality})} \label{sec:optimality}
In this section we show that GMA is optimal according to \labelcref{req-optimality}, which means that there is no better local (\labelcref{req-locality}) algorithm that does not over-allocate any edge or interface pair (\labelcref{req-no-overuse}).
As every invocation of a local algorithm is only based on the nodes of one path, and is oblivious of all the other nodes of the graph, in order to prevent overuse the algorithm has to consider all possible graphs containing this path.
This insight is central for the proof of optimality and is formalized in the following lemma:

\begin{lem}
    For every allocation graph and every one of its paths $\pi$, there exists another allocation graph that contains a path with the same sequence of allocation matrices, where the \pairalloc{} $\T{x}{i}{j}$ of some on-path node $x$ is fully utilized (there is no available resource left) if there is a \gma{} allocation on every path containing $(x, i^{x}, j^{x})$ in this new graph.
  \label{le:other_graph_full_utilization}
\end{lem}

\begin{myproof}
  Let $\pi$ be an arbitrary path of an arbitrary allocation graph, and let $x$ be the index for which \cref{eq:gma} is minimized.
  We construct a new allocation graph around $\pi$ as follows:
  \begin{itemize}
    \item Remove all the nodes that are not part of $\pi$.
    \item Keep the on-path nodes, their interfaces, and their allocation matrices as they are.
    \item For every node, create identical copies of the node for each of its occurrences on the path (multiple copies, in case the path contains loops) and only keep the edges to the previous and subsequent node on the path.
    \item For all these on-path nodes, attach new nodes to the non-local interfaces that are not already part of $\pi$. Those new nodes only have one local and one non-local interface (the interface through which they are attached to the on-path node).
    \item For every node $k \in \{1, \dots, x-1\}$ and each of its interfaces $\widetilde{i}$ to which a new node was attached, the \pairalloc{} (from its local to its non-local interface) of the new node is set to $\divr{k}{\widetilde{i}}$. This implies that also the divergent (at the local interface) and the convergent (at the non-local interface) of the new node are equal to $\divr{k}{\widetilde{i}}$.
    \item For $x$, the newly attached nodes can have arbitrary allocation-matrix entries.
    \item For every node~$k \in \{x+1, \dots, \ell\}$ and each of its interfaces $\widetilde{j}$ to which a new node was attached, the \pairalloc{} (from its non-local to its local interface) of the new node is set to $\conv{k}{\widetilde{j}}$. This implies that also the divergent (at the non-local interface) and the convergent (at the local interface) of the new node are equal to $\conv{k}{\widetilde{j}}$.
  \end{itemize}
  Given that there is a GMA allocation on every possible path (in our new graph) going through $(i^{x}, j^{x})$, we want to show that $\T{x}{i}{j}$ is fully utilized.
  We characterize all possible paths for three cases:
  If $1 < x < \ell$ (case 1), a path starts at a local interface of some node $k \leq x-1$ or at the local interface of some of its attached nodes, and ends at a local interface of some node $m \geq x+1$ or at the local interface of some of its attached nodes.
  If $x = 1$ (case 2), every path starts at the local interface of $x$, and ends at a local interface of some node $k \geq 2$ or at the local interface of some of its attached nodes.
  If $x = \ell$ (case 3), every path starts at a local interface of some node $k \leq \ell-1$ or at the local interface of some of its attached nodes, and ends at the local interface of node $\ell$.

  \paragraphProofNobreak{Case 1:}
  We use the following notation in order to simplify our proof:
  \begin{nospaceflalign}
    \au{u} = \frac{\T{u}{i}{j}}{\conv{u}{j}},~ \bu{u} = \frac{\T{u}{i}{j}}{\divr{u}{i}}
  \end{nospaceflalign}
  Let $R_u$ be the sum of all allocations of all the nodes $k \in \{1, \dots, x-1\}$ starting either at a local interface or at the local interface of some of its attached nodes, and ending either at a local interface of node $u$ or at the local interface of some of its attached nodes, divided by~$\T{x}{i}{j}$.
  Thus, we need to prove
  \begin{nospaceflalign}
    \T{x}{i}{j}\cdot\sum_{u=x+1}^{\ell} R_u = \T{x}{i}{j}
    \quad\Leftrightarrow\quad
    \sum_{u=x+1}^{\ell} R_u = 1.
  \end{nospaceflalign}
  We formulate two lemmas, which are proven in \appref{D}:
  \begin{lem}
    For $a_1, \dots, a_x > 0$: $\prod_{i=1}^{x} a_i + \sum_{k=1}^{x}\left((1-a_k)\cdot\prod_{i=k+1}^{x} a_i\right) = 1$.
    \label{le:cancels-to-one}
  \end{lem}
  \begin{lem}
    $R_\ell =\prod_{k=x+1}^{\ell -1}\bu{k}$ and $R_u =(\prod_{k=x+1}^{u-1}\bu{k})\cdot(1-\bu{u})$ (for $x+1 \leq u \leq \ell -1$).
    \label{le:Rl-Ru}
  \end{lem}
  These lemmas immediately imply our proof goal:
    \begin{nospaceflalign}
      \smashoperator[r]{\sum_{u=x+1}^{\ell}} R_u = \smashoperator{\sum_{u=x+1}^{\ell-1}} R_u + R_\ell
        &= \sum_{u=x+1}^{\ell -1}\smashoperator[r]{\prod_{k=x+1}^{u-1}} \bu{k}\cdot(1-\bu{u})+ \smashoperator{\prod_{k=x+1}^{\ell -1}} \bu{k}
        = 1.
    \end{nospaceflalign}
  \medskip

  \paragraphProofNobreak{Case 2+3:}
  The proofs follow a simplified structure of the proof of case 1.
\end{myproof}

\begin{thm}
  GMA is Pareto optimal among all algorithms in the sense of \labelcref{req-optimality}.
\end{thm}

\begin{myproof}
  This follows directly from \cref{le:other_graph_full_utilization}:
  for a given path (nodes with their associated allocation matrices) there always exists a graph containing that path, where increasing the allocation calculated by GMA will cause overuse, which can only be prevented by decreasing allocations on other paths.
\end{myproof}

\section{GMA provides meaningful allocations}
\label{sec:eval-allocations}
A potential limitation of \gma{} is the size of the allocations it provides.
We proved that \gma{}'s \pallocs{} are small enough that, even if all the paths have an allocation, no over-allocation occurs.
In this section we show that \gma{}'s \palloc{s} are still large enough to satisfy the requirements of the critical applications that motivate this work (details in \appref{A}). We do this by simulating \gma{} on random graphs, thereby exploring the trade-offs between graph topology and the resulting \gma{} allocation sizes.

\subsection{Simulation setup}

\paragraph{Graph topology}
We use the well-known Barabási--Albert random graph model to generate allocation graphs~\cite{Barabasi1999}. This  algorithm is designed to produce scale-free random graphs, which are found to well approximate real-life technological networks~\cite{Broido2019}.

At the topological level, the size of a \gma{} allocation for some path depends on (i) the degree of the nodes on the path, as
it determines the size of the allocation matrix, (ii) the length of the path, since \cref{eq:gma} contains an iterative product on each node on the path, and (iii) the capacity of each on-path edge (discussed in the next paragraph).
We aggregate the first two metrics at the graph level by considering the average node degree and the diameter of the graph, i.e., the length of the longest path.%
\footnote{These two factors are closely related with each other and to the number of nodes in the graph: keeping the number of nodes fixed, a graph with higher average node degree will inevitably have smaller diameter.}
Therefore, we generate $275$ random graphs for our simulations, with \num{8} to $2048$ nodes, varying average degree and diameter.
Additional details on graph generation can be found in \appref{G}.

\paragraph{Resources and \Allocmatrix{s}}
\label{sec:resource_and_alloc}
In the simulations, we model the varying bandwidth of real-world network links by assigning different capacities to the edges of graphs.
To assign capacity to edges based on a \emph{degree--gravity} model: the capacity of a (directed) edge is selected proportionally to the product of the degrees of its adjacent nodes~\cite{Saino2013}.
We discretize these values to $10$ different levels from $40$ to $400$.
This choice is motivated by real networks, where more connected nodes also tend to have higher forwarding capabilities.

Based on these edge capacities, we then create the \allocmatrix{s}.
Although each node might have different policies, simulating those policies for the nodes introduces many additional degrees of complexity, beyond the scope of this evaluation.
Therefore, we assume a simple \emph{proportional sharing} policy to construct an \allocmatrix{}, which we obtain by performing the following three steps for each node $k$ and all its interfaces $i$ and $j$:
(i) $\T{k}{i}{j} \gets \ifcap{k}{i}$, while for the local interface $\bot$, $\T{k}{\bot}{j}, \T{k}{i}{\bot} \gets \max_{i} \{\ifcap{k}{i}\}$;
(ii) $\T{k}{i}{j}\gets\T{k}{i}{j}\cdot\ifcap{k}{j}/\conv{k}{j}$;
(iii) if $\divr{k}{i} > \ifcap{k}{j}$, then $\T{k}{i}{j}\gets\T{k}{i}{j}\cdot\ifcap{k}{i}/\divr{k}{i}$.

\paragraph{Path selection}
In this simulation, the goal is to create \palloc{s} between every pair of nodes.
Motivated again by networking practice, we consider allocations made on $k$-shortest paths, with $k \in \{1, 2, 3\}$.
For $k=1$, we create allocations on the single-shortest path for every pair of nodes.
However, \gma{} can compute an allocation for \emph{any} path in the graph. Therefore, if two nodes are able to use multiple paths simultaneously, the total allocation for the pair is the aggregate of the allocations on the individual paths. We then create allocations on the $2$- and $3$-shortest paths for every pair of nodes, and evaluate the advantage that multipath communication can provide.

\paragraph{Metrics: \acover{$\boldsymbol\alpha$}}
Given a source node, the size of the \gma{} allocations to different destination nodes can vary greatly, and computing average statistics does not reflect the binary nature of critical application requirements: either the allocation exceeds the minimum usability threshold, or the allocation is not useful (see \appref{A} for details).

Therefore, we introduce a new metric to aggregate this information and compare the effectiveness of \gma{} across different topologies, called \acover{}.
Given a source node in a graph and a path selection strategy, the node's \acover{} is the fraction of destination nodes to which the sum of the \pallocs{} computed over the available paths is more than $\alpha$. Therefore, \acover{} captures the size of the sub-graph with which the source node can communicate using an adequately-sized \gma{} allocation.
For example, a node with a \acover{$10^{-4}$} of $0.7$ can reach \SI{70}{\percent} of the nodes in the graphs with an allocation of at least $10^{-4}$.
Naturally, higher values of \acover{} are better.
We define the median \acover{} of a graph as the median of the \acover{}s of its nodes (and similarly for minimum and maximum).
While different applications will require different values of $\alpha$, we use a \acover{$10^{-4}$} in all simulations.
Again, this is motivated by practical considerations: if we set \num{1}~unit of resource~=~\SI{1}{\gbps}, $10^{-4}$~units correspond to \SI{100}{\kbps}. The applications that motivate this work, such as blockchains and inter-bank transaction clearing, can comfortably operate within this boundary.

\begin{figure}[!t]
  \centering
    \begin{minipage}[t]{\textwidth}
        \centering
      \includegraphics[width=0.6\linewidth]{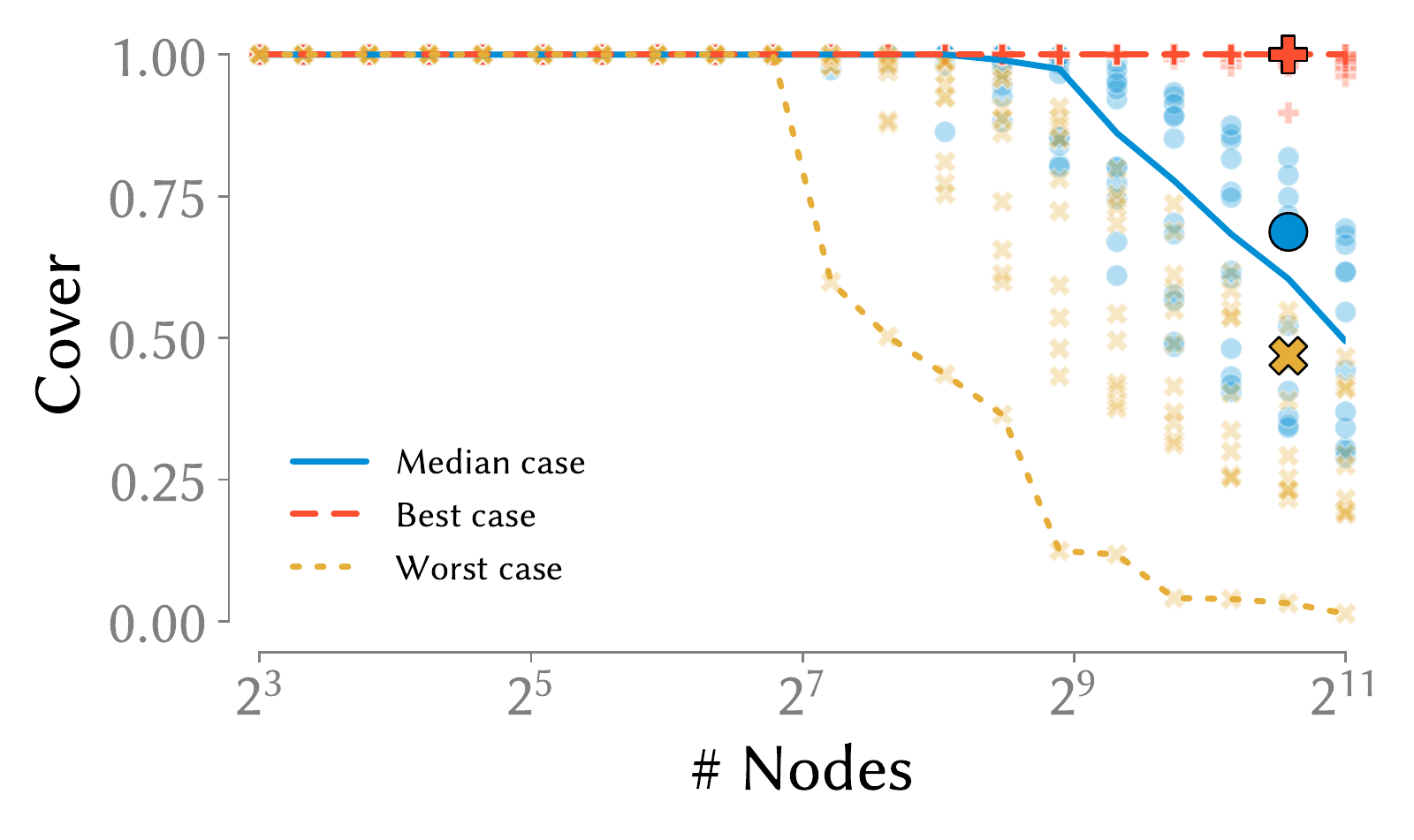}
    \caption{\textbf{Minimum, maximum, and median single-path \protect\acover{$\boldsymbol{10^{-4}}$}.}
    The highlighted markers show the max \ding{58}, median \ding{108}, and min \ding{54} cover for one specific graph (which is further analyzed in \cref{fig:barabasi_cover_diameter,fig:single_graph} in \cref{app:G}).}
    \label{fig:barabasi_cover_nodes}
  \end{minipage}
  \end{figure}

\subsection{Results}
For each of the generated graphs, \cref{fig:barabasi_cover_nodes} relates its minimum, maximum, and median \acover{$10^{-4}$} to the number of nodes, where we used the single shortest path selection scheme.
We see that all graphs have a median cover in the upper \SI{50}{\percent} range , while the minimum cover decreases to just a few percent for graphs with a high number of nodes. Graphs with lower median cover are the ones that have low or high diameter, as \appfigref{5}{G} shows. This confirms the observation that large allocation matrices (low diameter) or long paths (high diameter) decrease the size of allocations.
Further, in all graphs, we find at least one node with cover greater than \SI{89}{\percent}, and observe
that the cover increases with the degree of the nodes: central nodes have therefore better cover, an important property in practical applications. An example is shown in \appfigref{5}{G}.

\Appfigref{3}{G} shows the improvement in the median cover of the graphs when using the 2- or 3-shortest path selection schemes in place of of the single shortest path selection scheme.
We see that the returns for using additional paths are high, reaching over \SI{120}{\percent} increase over single-path cover when using three paths instead of one.
Graphs with lower number of nodes benefit less from the additional paths, as many already achieve perfect cover.
A higher $k$ could further increase the cover, but this exploration is left to future work.

\section{Related work}\label{sec:related}

\paragraph{Flow problems and algorithms}
A class of theoretical problems that are related to our path-allocation problem
are \emph{multi-commodity flow problems}, which have been studied extensively
since the 1950s~\cite{ford1958suggested}. The variant which is most closely
related to our setting is the \emph{maximum concurrent flow
problem}~\cite{Shahrokhi1990}, where fairness between different commodities is
taken into account, but the ratios are set by a central controller. All
variants differ from our \pad{} problem in that they (i) do not consider
independent nodes with their own properties and (ii) require a global knowledge
of the topology. They have thus been applied mostly to centrally controlled
networks~\cite{Chang2018}.

\paragraph{Resource allocation in networks}
Bandwidth guarantees were a central concept of virtual-circuit architectures
like ATM~\cite{ATM}. For today's IP-based Internet, bandwidth reservations have
been proposed in the Integrated Services (IntServ) architecture~\cite{RFC1633},
in which they are negotiated through the Resource Reservation Protocol
(RSVP)~\cite{RFC2205}. However, due to its high reliance on in-network state,
IntServ has never been widely adopted. Further, these systems do not specify
\emph{how much} bandwidth should be allocated to flows.
The Internet overwhelmingly relies on congestion
control~\cite{Jacobson1988,Kelly1998} as a distributed mechanism for bandwidth
allocation between flows, which provides no guarantees to the communication
partners and has no support to implement traffic policies. There exists a wide
range of traffic-engineering systems suitable to intra-domain contexts, such as
MPLS~\cite{RFC3031} with OSPF-TE~\cite{RFC3630} and RSVP-TE~\cite{RFC3209} or
SDN-based solutions~\cite{Shu2016}. However, in contrast to \gma{}, which
supports autonomous nodes, all these systems require a central controller.

\section{Discussion and Conclusion}\label{sec:conclusion}

In this paper, we revisit an old networking and distributed-systems
problem---how to allocate resources in a network of independent nodes when no
central controller is available. After introducing the formalism of allocation
graphs, in which each node is associated with \emph{local} allocations based on
available resources and policies, we ask a novel question: can an algorithm
compute resource allocations for all paths in an allocation graph, without
causing over-allocation, and relying only on local information? This is the
foundation of the \pad{} problem. We answer with our \emph{global myopic
allocation} (GMA) algorithm, showing how these local decisions give rise to
meaningful and sustainable \emph{global} allocations.  Further, we prove that
these allocations are Pareto-optimal, and therefore cannot be trivially
improved.

\paragraph{Relevance to networking} The allocations calculated through GMA are
static and depend only on the policies of on-path nodes; in particular, they
are independent of other allocations and resource demands. They thus provide
strong minimal resource guarantees that are valid under all networking
conditions and are particularly relevant for applications where centralized
solutions based on dedicated network infrastructure are too expensive or
inherently impossible. By their very nature, these guaranteed allocations are
smaller than what can be achieved through dynamic resource-allocation systems.
However, our simulations show that, even under conservative assumptions, \gma{}
provides sufficient communication bandwidth to virtually all pairs of nodes in
small to medium-sized networks. Thus, \gma{}-based allocations with strong
availability guarantees could \emph{complement} other systems with higher
network utilization but weaker guarantees, such as best-effort traffic.

\paragraph{Future work} The novel results on graph resource allocation
presented in this paper open many new and exciting avenues for future research,
both theoretical and applied. First of all, this paper did not explore the
\emph{fairness} implications of \gma{} allocations. The properties of
monotonicity and Pareto-optimality, along with the proportional use of
\pairalloc{s} in the computation, point towards a strong \emph{neighbor-based}
fairness notion. We leave the analysis of such a notion to future work.
Second, we see great potential for further research on \pad{} algorithms.
For instance, Pareto optimality does not satisfy the question of whether \gma{}
is optimal in a global sense, i.e., whether it maximizes a function over all
\pallocs{}---their sum, for example. The discovery of globally optimal \pad{}
algorithms could lead to interesting theoretical advancements, with profound
practical implications.

Finally, in this paper we have discussed how allocations can be \emph{computed}
in a distributed setting. This is orthogonal to the development of specific
protocols necessary to communicate and authenticate necessary information and
enforce the allocations. Future research could focus on the development of such
a protocol and investigate its interplay with other networking paradigms like
best-effort traffic and congestion control.

\section*{Acknowledgments}

We would like to thank Mohsen Ghaffari for the illuminating discussions;
Tobias Klenze, Simon Scherrer, Stefan Schmid, and Joel Wanner for their feedback on the manuscript;
and the anonymous reviewers for their
insightful comments.

%
%
%
\clearpage
\appendix
\newpage

\newcommand{\Fac}{\ensuremath{F}}
\newcommand{\FacNew}{\ensuremath{\tilde{\Fac}}}

\section{Critical networking applications}
\label{app:crit_app}
\label{app:A}

For many \emph{critical} applications, reliability, security, and
scalability of communication systems are of paramount importance.
These application  require relatively low traffic volumes, but availability has to be guaranteed at all times for these services to achieve their task.
We provide two examples of such applications.

The first is inter-bank transactions. The SWIFT financial messaging network is a prominent example in this market, as it handles transactions between its \num{11000} member institutions and accounts for half of global cross-border inter-bank transactions~\cite{SWIFT_FT}.
Despite the importance of these transactions for today's financial system, their actual bandwidth requirements are modest.
On an average day, SWIFT processes around \num{40} million messages in total~\cite{SWIFT_Stat}, which corresponds to fewer than \num{500} messages per second---globally.
Each transaction is encoded in an XML file of variable size, usually around a few kilobytes (estimate based on real-world examples of the XML-encoded ISO 20022 transaction message format~\cite{swift_message_definition}), resulting on an average load of less than \SI{1}{\mbps} between all \num{11000} institutions.

The Bitcoin network provides a second example. Each Bitcoin miner node needs to
run the consensus protocol in order to verify  the transaction that are being
committed to the blockchain. Today, the network processes 7 transactions per
second~\cite{Croman_2016}, with an average transaction size of \SI{500}{\byte},
and very rarely above \SI{1}{\kB}~\cite{BitcoinSize1,BitcoinSize2}.
This directly translates
to modest bandwidth requirements of less than \SI{100}{\kbps} per node.
However, delays or interruptions of communication can result in financial
loss~\cite{Apostolaki2017}. A further complication complication that arises in
blockchain networks is decentralization. As nodes are run by different---and
often untrusted---entities, centralized solutions are avoided as they introduce
a single point of failure. Even permissioned blockchains, like the Libra
network, impose node decentralization by design as a way to build
trust~\cite{Libra_whitepaper}.

In general, critical applications share these common traits: (i) the required traffic volumes are relatively small, less than \SI{100}{\kbps} per end-to-end communication, but (ii) connectivity has to be ensured at all times (availability), (iii) even in the presence of denial-of-service (DoS) attacks (security). Finally, (iv) the guarantees have to be extended to large networks, in many cases under the assumption of decentralized control.

\section{Cases in which down-scaling improves the allocation calculated by \cref{eq:gma_old}}
\label{app:scale-down}
\label{app:B}

\begin{lem}
  Let $\pi$ be an arbitrary path consisting of $\ell$ nodes, and let $k$ be one if its on-path nodes. If $1 < k < \ell$, scaling down its contributed values $(\divr{k}{i}, \T{k}{i}{j},$ $ \conv{k}{j})$ (without scaling down any values of other nodes) can only improve \atwo{\pi} if $\divr{k}{i} > \conv{k-1}{j}$ and $\conv{k}{j} > \divr{k+1}{i}$.\\ If $k = 1$ or $k = \ell$, scaling-down its hop values will never increase \atwo{\pi}.
  \label{le:scaling-down}
\end{lem}
\begin{myproof}~\\
  \paragraphProofNobreak{Case $1 < k < \ell$:}
  The contributed values of node $k$ are part of the following factor of \cref{eq:gma_old}:
  \begin{nospaceflalign}
    \Fac{}:=\frac{\T{k}{i}{j}}{\max\{\conv{k-1}{j}, \divr{k}{i} \} \cdot \max\{\conv{k}{j}, \divr{k+1}{i} \}}.
    \label{eq:gma-contributed-terms}
  \end{nospaceflalign}
  We write $\FacNew{}$ for the same factor after scaling down the values $(\divr{k}{i}, \T{k}{i}{j},$ $\conv{k}{j})$.
  \begin{subparagraphindent}
    \paragraphProof{Case $(\divr{k}{i} > \conv{k-1}{j}) \land (\conv{k}{j} > \divr{k+1}{i})$:}
    Scaling down the contributed values by $s := \conv{k-1}{j} / \divr{k}{i} < 1$ leads to
    \begin{nospaceflalign}
      \FacNew{} = \frac{s\cdot\T{k}{i}{j}}{s\cdot\divr{k}{i} \cdot \max\{s \cdot\conv{k}{j}, \divr{k+1}{i} \}} > \frac{\T{k}{i}{j}}{\divr{k}{i} \cdot \conv{k}{j}} = \Fac{}.
    \end{nospaceflalign}

    \paragraphProof{Case $(\divr{k}{i} > \conv{k-1}{j}) \land (\conv{k}{j} \leq \divr{k+1}{i})$:}
    Scaling down the contributed values by $s$ where $\conv{k-1}{j} / \divr{k}{i} \leq s < 1$ has no impact on \cref{eq:gma-contributed-terms}:
    \begin{nospaceflalign}
      \FacNew{} = \frac{s\cdot\T{k}{i}{j}}{s\cdot\divr{k}{i} \cdot \max\{s \cdot\conv{k}{j}, \divr{k+1}{i} \}} = \frac{\T{k}{i}{j}}{\divr{k}{i} \cdot \divr{k+1}{i}} = \Fac{}.
    \end{nospaceflalign}
    Any further down-scaling only decreases the allocation, as shown in the last case.

    \paragraphProof{Case $(\divr{k}{i} \leq \conv{k-1}{j}) \land (\conv{k}{j} > \divr{k+1}{i})$:}
    The proof follows the same structure as in the previous case.

    \paragraphProof{Case $(\divr{k}{i} \leq \conv{k-1}{j}) \land (\conv{k}{j} \leq \divr{k+1}{i})$:}
    Scaling down the contributed values by any factor $s < 1$ leads to
    \begin{subequations}%
    \begin{nospaceflalign}
      \FacNew{} &=\frac{s\cdot\T{k}{i}{j}}{\max\{\conv{k-1}{j}, s\cdot\divr{k}{i} \} \cdot \max\{s\cdot\conv{k}{j}, \divr{k+1}{i} \}}
      \\
      &= \frac{s\cdot\T{k}{i}{j}}{\conv{k-1}{j} \cdot \divr{k+1}{i}}
      < \frac{\T{k}{i}{j}}{\conv{k-1}{j} \cdot \divr{k+1}{i}} = \Fac{}.
    \end{nospaceflalign}
    \end{subequations}

  \end{subparagraphindent}

  \paragraphProofNobreak{Case $k = \ell$:}
  The contributed values of node $\ell$ are part of the following factor of \cref{eq:gma_old}:
  \begin{nospaceflalign}
    \frac{\T{\ell}{i}{j}}{\max\{\conv{\ell-1}{j}, \divr{\ell}{i} \}}
    \label{eq:gma-contributed-terms-L}
  \end{nospaceflalign}
  Scaling down the contributed values by any factor $s < 1$ modifies \cref{eq:gma-contributed-terms-L} to
  \begin{subequations}
  \begin{nospaceflalign}
    \frac{s\cdot\T{\ell}{i}{j}}{\max\{\conv{\ell-1}{j}, s\cdot\divr{\ell}{i} \}}
    &= \frac{\T{\ell}{i}{j}}{\max\{\frac{1}{s}\cdot\conv{\ell-1}{j}, \divr{\ell}{i} \}} \\
    &\leq \frac{\T{\ell}{i}{j}}{\max\{\conv{\ell-1}{j}, \divr{\ell}{i} \}}.
  \end{nospaceflalign}
  \end{subequations}

  \paragraphProofNobreak{Case $k = 1$:}
  The proof follows the same structure as in the case $k=\ell$.
\end{myproof}

\section{Equivalence of recursive and direct \gma{} formulas}
\label{app:proof_equivalent_formula}
\label{app:C}

\begin{lem}
  \Cref{eq:gma-f} is equivalent to \cref{eq:gma}.
    \label{le:gma-equivalence}
\end{lem}

\begin{myproof}
  We prove \cref{le:gma-equivalence} by induction over the path length $\ell$.\medskip
  \paragraphProof{\\Base case ($\ell=1$):}
  \noindent Because $f^{(1)} = 1$, we get $\T{1}{i}{j} = f^{(1)}\cdot\T{1}{i}{j}$. \medskip

  \paragraphProofNobreak{Inductive step:}
  \begin{subparagraphindent}
    \paragraphProofNobreak{Induction hypothesis:}\\
    For a particular~$\ell$:
    \begin{nospaceflalign*}
       &&\frac{f^{(\ell)}\cdot\prod_{k=1}^{\ell} \T{k}{i}{j}}{\prod_{k=2}^{\ell} \conv{k-1}{j}}
        = \min_{0\leq x\leq \ell} \biggl(\prod_{k=1}^{x-1} \frac{\T{k}{i}{j}}{\conv{k}{j}}\cdot \T{x}{i}{j} \cdot \prod_{k=x+1}^{\ell} \frac{\T{k}{i}{j}}{\divr{k}{i}} \biggr)
    \end{nospaceflalign*}
    \paragraphProofNobreak{To show:}
    \begin{nospaceflalign*}
      \indent\text{}~ &&\frac{f^{(\ell +1)}\cdot\prod_{k=1}^{\ell +1} \T{k}{i}{j}}{\prod_{k=2}^{\ell +1} \conv{k-1}{j}}
        = \min_{0\leq x\leq \ell+1} \biggl(\prod_{k=1}^{x-1} \frac{\T{k}{i}{j}}{\conv{k}{j}}\cdot \T{x}{i}{j} \cdot \prod_{k=x+1}^{\ell +1} \frac{\T{k}{i}{j}}{\divr{k}{i}} \biggr)
    \end{nospaceflalign*}
  \end{subparagraphindent} \smallskip

  \subparagraphProofNobreak{Proof:} \medskip
  \allowdisplaybreaks
  \begin{subequations}
    \begin{nospaceflalign}
      \indent&\frac{f^{(\ell +1)}\cdot\prod_{k=1}^{\ell +1} \T{k}{i}{j}}{\prod_{k=2}^{\ell +1} \conv{k-1}{j}}
        = \min \biggl(1,~~\frac{\conv{\ell}{j}\cdot f^{(\ell)}}{\divr{\ell +1}{i}}\biggr)\cdot \frac{\prod_{k=1}^{\ell +1} \T{k}{i}{j}}{\prod_{k=2}^{\ell +1} \conv{k-1}{j}}
          \label{eq:gma-equivalence-f-out} \\
        &= \min \biggl(\frac{\prod_{k=1}^{\ell +1} \T{k}{i}{j}}{\prod_{k=2}^{\ell +1} \conv{k-1}{j}},~~\frac{\T{\ell +1}{i}{j}}{\divr{\ell +1}{i}}\cdot f^{(\ell)}\cdot\frac{\prod_{k=1}^{\ell} \T{k}{i}{j}}{\prod_{k=2}^{\ell} \conv{k-1}{j}} \biggr)
          \label{eq:gma-equivalence-into-min} \\
        &= \min \biggl(\frac{\prod_{k=1}^{\ell +1} \T{k}{i}{j}}{\prod_{k=2}^{\ell +1} \conv{k-1}{j}},~\min_{0\leq x\leq \ell} \biggl(\prod_{k=1}^{x-1} \frac{\T{k}{i}{j}}{\conv{k}{j}}\cdot \T{x}{i}{j} \cdot \prod_{k=x+1}^{\ell +1} \frac{\T{k}{i}{j}}{\divr{k}{i}} \biggr) \biggr)
          \label{eq:gma-equivalence-equivalence} \\
        &= \min_{0\leq x\leq \ell+1} \biggl(\prod_{k=1}^{x-1} \frac{\T{k}{i}{j}}{\conv{k}{j}}\cdot \T{x}{i}{j} \cdot \prod_{k=x+1}^{\ell +1} \frac{\T{k}{i}{j}}{\divr{k}{i}} \biggr)
          \label{eq:gma-equivalence-final}
    \end{nospaceflalign}
  \end{subequations}
  \begin{subparagraphindent}
    In the first step we applied the definition of $f$ from \cref{eq:gma-f-recursive}. To get \cref{eq:gma-equivalence-into-min} we moved the rightmost factor into the $\min$ term, and in the following step used the induction hypothesis. The last equation follows from $\displaystyle\min_{1\leq x\leq \ell +1}\left(g(x)\right) = \min\big(g(\ell +1), \min_{1\leq x\leq \ell}\left(g(x)\right) \big)$, which holds for any function $g$. \qedhere
  \end{subparagraphindent}
\end{myproof}

\section{Lemmas used in the proof of optimality}
\label{app:lemmas_opt}
\label{app:D}
\allowdisplaybreaks

We first formulate some additional lemmas and then prove \cref{le:cancels-to-one,le:Rl-Ru} used in \cref{sec:no-overuse}.
To simplify the notation, we drop the nodes in the paths in this section.

\subsection{Auxiliary lemmas}
\label{sec:lemmas_opt.auxiliary}

In the following lemmas we consider an arbitrary path $\pi = [ (i^1, j^1), (i^2, j^2),\dots,$ $(i^{\ell}, j^{\ell}) ]$ and denote the index for which \cref{eq:gma} is minimized as \xmin{}.

\begin{lem}
  If $\xmin \geq 3$, then the \gma{} allocation for the path $\widetilde{\pi} = [(\widetilde{i^2}, j^2),\dots,$ $(i^{\ell}, j^{\ell}) ]$ beginning at some interface of node 2 is still minimized at node~\xmin.
  \label{le:left-cut-preserves-min-x}
\end{lem}

\begin{myproof}
  \begin{subequations}
    \begin{nospaceflalign}
        \xmin
        &= \argmin_{x} \left(\prod_{k=1}^{x-1} \frac{\T{k}{i}{j}}{\conv{k}{j}}\cdot \T{x}{i}{j} \cdot \prod_{k=x+1}^{\ell} \frac{\T{k}{i}{j}}{\divr{k}{i}} \right)
          \label{eq:left-cut-start} \\
        &= \argmin_{x} \left(\prod_{k=1}^{x-1} \frac{1}{\conv{k}{j}}\cdot \prod_{k=x+1}^{\ell} \frac{1}{\divr{k}{i}} \right)
          \label{eq:left-cut-no-M} \\
        &= \argmin_{x} \left(\prod_{k=2}^{x-1} \frac{1}{\conv{k}{j}}\cdot \prod_{k=x+1}^{\ell} \frac{1}{\divr{k}{i}} \right)
          \label{eq:left-cut-k2} \\
        &= \argmin_{x} \left(\frac{\T{2}{\widetilde{i}}{j}}{\conv{2}{j}}\cdot\prod_{k=3}^{x-1} \frac{\T{k}{i}{j}}{\conv{k}{j}}\cdot \T{x}{i}{j} \cdot\prod_{k=x+1}^{\ell} \frac{\T{k}{i}{j}}{\divr{k}{i}} \right)
          \label{eq:left-cut-final}
    \end{nospaceflalign}
  \end{subequations}
  In \cref{eq:left-cut-no-M,eq:left-cut-final} we used the fact that the \matrixentry{s} do not contribute to the argument of the minimum (they are constant among all the possible values for $x$). This is also true for the convergent of node $1$ (because $x \geq 3$), which is used in \cref{eq:left-cut-k2}.
\end{myproof}

\begin{lem}
  If $\xmin{} \leq \ell -2$, then the \gma{} allocation for the path $\widetilde{\pi} = [(i^1, j^1),$ $(i^2, j^2), \dots, (i^{\ell -1}, \widetilde{j^{\ell -1}}) ]$ ending at some interface of node $\ell -1$ is still minimized at node~\xmin{}.
  \label{le:right-cut-preserves-min-x}
\end{lem}

\begin{myproof}
  The proof follows the same structure as the proof of \cref{le:left-cut-preserves-min-x}.
\end{myproof}

\begin{lem}
  When extending the path on a non-local interface $\widetilde{i^1}$ of node 1 with some node 0 that only consists of a local and a non-local interface to $\widetilde{\pi} = [ (i^0, j^0), (\widetilde{i^1}, j^1),\dots, (i^{\ell}, j^{\ell}) ]$, and given that $\T{0}{i}{j} = \conv{0}{j} = \divr{1}{\widetilde{i}}$, then the resulting allocation will be independent of the \allocmatrix{} of node 0 and will still be minimized at node \xmin{}.
  \label{le:left-extend-preserves-alloc}
\end{lem}

\begin{myproof}
  \newcommand\myeq{~\mathrel{\stackrel{\makebox[0pt]{\mbox{\normalfont\tiny by~def}}}{=}}~}
  We define
  \begin{subequations}
  \begin{nospaceflalign}
      g_1(x) &:= \prod_{k=1}^\ell \T{k}{i}{j}\cdot \prod_{k=1}^{x-1} \frac{1}{\conv{k}{j}}\cdot \prod_{k=x+1}^{\ell} \frac{1}{\divr{k}{i}},
      \\
      g_0(x) &:= \T{0}{i}{j}\cdot\T{1}{\widetilde{i}}{j}\cdot\prod_{k=2}^\ell \T{k}{i}{j}\cdot \prod_{k=0}^{x-1} \frac{1}{\conv{k}{j}}\cdot \prod_{k=x+1}^{\ell} \frac{1}{\divr{k}{i}}.\label{eq:def.g0}
  \end{nospaceflalign}
  \end{subequations}
  We see that
  \begin{nospaceflalign}
    g_1(\xmin{})
    &= \gmaG{\pi}~\myeq~\prod_{k=1}^\ell \T{k}{i}{j}\cdot \min_{1\leq x\leq \ell} \biggl(\prod_{k=1}^{x-1} \frac{1}{\conv{k}{j}}\cdot \prod_{k=x+1}^{\ell} \frac{1}{\divr{k}{i}}\biggr),
      \label{eq:extending-start}
  \end{nospaceflalign}
  because \xmin{} is the argument of the minimum of $\gmaG{\pi}$. By multiplying the terms on both sides of the equation with $\frac{\T{0}{i}{j}\cdot \T{1}{\widetilde{i}}{j}}{\conv{0}{j}\cdot \T{1}{i}{j}}$, we get
  \begin{subequations}
    \begin{nospaceflalign}
    g_0(\xmin{})
    &= \T{0}{i}{j}\cdot\T{1}{\widetilde{i}}{j}\cdot\prod_{k=2}^\ell \T{k}{i}{j}\cdot \min_{1\leq x\leq \ell} \biggl(\prod_{k=0}^{x-1} \frac{1}{\conv{k}{j}}\cdot \prod_{k=x+1}^{\ell} \frac{1}{\divr{k}{i}}\biggr)
      \label{eq:extending-multiplied} \\
    %
    &= \T{0}{i}{j}\cdot\T{1}{\widetilde{i}}{j}\cdot\prod_{k=2}^\ell \T{k}{i}{j}\cdot \min_{0\leq x\leq \ell} \biggl(\prod_{k=0}^{x-1} \frac{1}{\conv{k}{j}}\cdot \prod_{k=x+1}^{\ell} \frac{1}{\divr{k}{i}}\biggr) \label{eq:extending-final} \\
    &\myeq~\gmaG{\widetilde{\pi}}.
\end{nospaceflalign}
\end{subequations}
  \Cref{eq:extending-final} shows that $\gmaG{\widetilde{\pi}}$ is still minimized at node \xmin{}; it follows from the inequality
  \begin{subequations}
  \begin{nospaceflalign}
    g_0(\xmin{})
      &\leq\frac{\T{0}{i}{j}}{\conv{0}{j}}\cdot\T{1}{\widetilde{i}}{j}\cdot\prod_{k=2}^\ell \T{k}{i}{j}\cdot \prod_{k=2}^{\ell} \frac{1}{\divr{k}{i}} \\
      &=\T{0}{i}{j}\cdot\T{1}{\widetilde{i}}{j}\cdot\prod_{k=2}^\ell \T{k}{i}{j}\cdot \prod_{k=1}^{\ell} \frac{1}{\divr{k}{i}}.
  \end{nospaceflalign}
  \end{subequations}
  Here we first used that, as~\xmin{} minimizes the right side of \cref{eq:extending-multiplied}, $g_0(\xmin{})$ is at most as high as the expression of the minimum for index~$1$. The second step follows from $\conv{0}{j} = \divr{1}{\widetilde{i}}$.
  Note that the resulting allocation $g_0(x)$ on path $\widetilde{\pi}$ is independent of node~$0$, because $\T{0}{i}{j}~=~\conv{0}{j}$, meaning that those terms cancel each other out, see \cref{eq:def.g0}.
\end{myproof}

\begin{lem}
  When extending the path on a non-local interface $\widetilde{j^\ell}$ of node $\ell$ with some node $\ell +1$ that only consists of a local and a non-local interface to $\widetilde{\pi} = [ (i^1, j^1),\dots, (i^{\ell}, \widetilde{j^{\ell}}), (i^{\ell +1}, j^{\ell +1}) ]$, and given that $\T{\ell +1}{i}{j} = \conv{\ell +1}{i} = \divr{\ell}{\widetilde{j}}$, then the resulting allocation will be independent of the \allocmatrix{} of node $\ell +1$ and will still be minimized at node \xmin{}.
  \label{le:right-extend-preserves-alloc}
\end{lem}

\begin{myproof}
  The proof follows the same structure as the proof of \cref{le:left-extend-preserves-alloc}.
\end{myproof}

\subsection{Lemmas used in main text}
\label{sec:lemmas_opt.core}


\paragraph{\cref{le:cancels-to-one}}
For $a_1, \dots, a_x > 0$ it holds that
\begin{equation}
    \prod_{i=1}^{x} a_i + \sum_{k=1}^{x}\left((1-a_k)\cdot\prod_{i=k+1}^{x} a_i\right) = 1.
\end{equation}

\begin{myproof}
  We do the proof by induction. \medskip
  \paragraphProof{\\Base case ($x=1$):}
  \noindent $a_1 + (1-a_1) = 1$
  \smallskip

  \paragraphProofNobreak{Inductive step:}
  \begin{subparagraphindent}
    \paragraphProof{Induction hypothesis:}
    $\prod_{i=1}^{x}a_i + \sum_{k=1}^{x}\left((1-a_k)\cdot\prod_{i=k+1}^{x}a_i\right) = 1$.\smallskip \\
    \paragraphProof{To show:}
    $\prod_{i=1}^{x+1}a_i + \sum_{k=1}^{x+1}\left((1-a_k)\cdot\prod_{i=k+1}^{x+1}a_i\right) = 1$.
  \end{subparagraphindent} \smallskip
  \subparagraphProofNobreak{Proof:}
  \allowdisplaybreaks
  \begin{subequations}
    \begin{nospaceflalign}[0pt]
    &\prod_{i=1}^{x+1}a_i + \sum_{k=1}^{x+1}\left((1-a_k)\cdot\prod_{i=k+1}^{x+1}a_i\right)\\
      &= a_{x+1}\cdot\prod_{i=1}^{x}a_i + a_{x+1}\cdot\sum_{k=1}^{x}\left((1-a_k)\cdot\prod_{i=k}^{x+1}a_i\right)+ (1-a_{x+1}) \\
      &= a_{x+1}\cdot 1+ (1-a_{x+1})
        \label{eq:cancels-to-one-simplified}
      = 1
    \end{nospaceflalign}
  \end{subequations}
  In \cref{eq:cancels-to-one-simplified} we used the induction hypothesis.
\end{myproof}

\paragraph{\cref{le:Rl-Ru}}
We defined $x$ as the index for which \cref{eq:gma} is minimized and assume $1 < x < \ell$. We defined $R_u$ as the sum of all allocations of all the nodes $k \in \{1, \dots, x-1\}$ starting either at a local interface or at the local interface of some of its attached nodes, and ending either at a local interface of node $u$ or at the local interface of some of its attached nodes, divided by~$\T{x}{i}{j}$.

Then, it holds that
\begin{subequations}
    \begin{nospaceflalign}
        R_\ell &=\prod_{k=x+1}^{\ell -1}\bu{k},\\
        R_u &= (\prod_{k=x+1}^{u-1}\bu{k})\cdot(1-\bu{u}) \qquad \text{(for $x+1 \leq u \leq \ell -1$)}.
    \end{nospaceflalign}
\end{subequations}

\begin{myproof}
  Let \IF{u} be the set of interfaces of node $u$. For each node we will use $\bot$ to refer to its local interface.
  $R_\ell$ consists of all allocations starting at the local interface of node~$1$ plus all the allocations starting at the local interface of some node that is attached to one of the nodes~$1$ to~$x-1$, where the allocations are ending either at a local interface of node~$\ell$ or at the local interface of some of its attached nodes:
  \begin{multline}
        \label{eq:R-l-start}
      R_\ell = \prod_{k=1}^{x-1}\au{k} \cdot\prod_{k=x+1}^{\ell -1}\bu{k} \cdot\biggl(\bu{\ell} + \sum_{\substack{t \in\IF{\ell}\\\setminus\{i^\ell, \bot\}}}\frac{\T{\ell}{i}{t}}{\divr{\ell}{i}}\biggr)  \\
      +\sum_{1 \leq p \leq x-1} \sum_{\substack{t \in\IF{p}\\\setminus\{i^p, j^p\}}}\frac{\T{p}{t}{j}}{\conv{p}{j}}\cdot\prod_{k=p+1}^{x-1}\au{k} \cdot\prod_{k=x+1}^{\ell -1}\bu{k} \cdot\biggl(\bu{\ell} + \sum_{\substack{t \in\IF{\ell}\\\setminus\{i^\ell, \bot\}}}\frac{\T{\ell}{i}{t}}{\divr{\ell}{i}}\biggr).
  \end{multline}
  Note that for all paths going through $(i^x, j^x)$, the argument of the minimum of \cref{eq:gma} is always the index~$x$: every such path can be constructed from the initial path by first dropping interface pairs at its origin and its end, and then extending the reduced path with the attached nodes. Both operations preserve $x$ as the argument of the minimum of \cref{eq:gma}, as shown in \cref{le:left-cut-preserves-min-x,le:right-cut-preserves-min-x,le:left-extend-preserves-alloc,le:right-extend-preserves-alloc}.
  Furthermore, the attached nodes do not have an influence on the \gma{} allocation, which is a consequence of \cref{le:left-extend-preserves-alloc,le:right-extend-preserves-alloc}.
  We observed that $\bu{\ell} + \sum_{t\in\IF{\ell}\setminus\{i^\ell,\bot\}}\frac{\T{\ell}{i}{t}}{\divr{\ell}{i}} = 1$ and obtain
  \begin{subequations}
    \begin{align}
      R_\ell&=\prod_{k=1}^{x-1}\au{k} \cdot\prod_{k=x+1}^{\ell -1}\bu{k}+\sum_{1 \leq p \leq x-1} \sum_{\substack{t \in\IF{p}\\\setminus\{i^p, j^p\}}}\frac{\T{p}{t}{j}}{\conv{p}{j}}\cdot\prod_{k=p+1}^{x-1}\au{k} \cdot\prod_{k=x+1}^{\ell -1}\bu{k}
        \label{eq:R-l-equal-one} \\
      &=\prod_{k=1}^{x-1}\au{k} \cdot\prod_{k=x+1}^{\ell -1}\bu{k}+\sum_{1 \leq p \leq x-1}(1-\au{p})\cdot\prod_{k=p+1}^{x-1}\au{k} \cdot\prod_{k=x+1}^{\ell -1}\bu{k}
        \label{eq:R-l-replace-a}\\
      &=\prod_{k=x+1}^{\ell -1}\bu{k},
        \label{R-l-final}
    \end{align}
  \end{subequations}
  where we used the observation that $\sum_{t \in\IF{p}\\\setminus\{i^p, j^p\}}\frac{\T{p}{t}{j}}{\conv{p}{j}} = 1-\au{p}$ in the step to \cref{eq:R-l-replace-a} and \cref{le:cancels-to-one} for the last step.

  With the same reasoning as above, we get, for $x+1 \leq u \leq \ell -1$,
  \begin{subequations}
    \begin{align}
    R_u
      &=\prod_{k=1}^{x-1}\au{k} \cdot\prod_{k=x+1}^{u-1}\bu{k} \cdot\biggl(\sum_{\substack{t \in\IF{u}\\-\{i^u, j^u\}}}\frac{\T{u}{i}{t}}{\divr{u}{i}}\biggr) \nonumber \\
      &\hphantom{{}={}}+\sum_{1 \leq p \leq x-1}\sum_{\substack{t \in\IF{p}\\\setminus\{i^p, j^p\}}}\frac{\T{p}{t}{j}}{\conv{p}{j}}\cdot\prod_{k=2}^{x-1}\au{k} \cdot\prod_{k=x+1}^{u-1}\bu{k} \cdot\biggl(\sum_{\substack{t \in\IF{u}\\\setminus\{i^u, j^u\}}}\frac{\T{u}{i}{t}}{\divr{u}{i}}\biggr) \\
      &=\prod_{k=1}^{x-1}\au{k} \cdot\prod_{k=x+1}^{u-1}\bu{k}\cdot(1-\bu{u}) \\
      &\hphantom{{}={}}+\sum_{\mathclap{1 \leq p \leq x-1}}(1-\au{p})\cdot\prod_{k=p+1}^{x-1}\au{k}\cdot\prod_{k=x+1}^{u-1}\bu{k}\cdot(1-\bu{u}) \\
      &=\left(\prod_{k=1}^{x-1}\au{k} \cdot\prod_{k=x+1}^{u-1}\bu{k}+\sum_{\mathclap{1 \leq p \leq x-1}}(1-\au{p})\cdot\prod_{k=p+1}^{x-1}\au{k}\cdot\prod_{k=x+1}^{u-1}\bu{k}\right)\cdot(1-\bu{u}) \\
      &=\left(\prod_{k=x+1}^{u-1}\bu{k}\right)\cdot(1-\bu{u}).
    \end{align}
  \end{subequations}
\end{myproof}

\section{Proofs of supplementary properties}
\label{app:supplementary}
\label{app:E}

\paragraph{Usability (\labelcref{req-usability})}
For every valid path, all the \pairalloc{s} used to calculate the allocation are positive by definition.
Moreover, convergents and divergents at each node contain the respective \pairalloc{} as part of the sum in \cref{eq:div_con}, and are therefore positive.
Every allocation is then positive, as it is a product of positive factors (\cref{eq:gma}).
\qed

\paragraph{Efficiency (\labelcref{req-efficiency})}
The polynomial complexity of \gma{} follows directly from \cref{eq:gma-f,eq:gma-f-recursive}. In fact, \gma{} has \emph{linear} complexity in the path length (assuming convergents and divergents are precomputed together with the \allocmatrix{s}).
\qed

\paragraph{Monotonicity (\labelcref{req-monotonicity})} \label{sec:monotonicity}
In the proof of monotonicity we will make use of the following lemma:
\begin{lem}
  If $a,b,\delta > 0$ and $a \leq b$, then it holds that $\frac{a+\delta}{b+\delta}\geq\frac{a}{b}$.
  \label{le:increase-by-delta}
\end{lem}

\begin{myproof}
  $\frac{a+\delta}{b+\delta} = \frac{a}{b}\cdot\frac{b\cdot(a+\delta)}{a\cdot(b+\delta)}
  =\frac{a}{b}\cdot\frac{ab + b\delta}{ab+a\delta}
  =\frac{a}{b}\cdot\left(1 + \frac{\delta(b-a)}{ab+a\delta}\right) \geq \frac{a}{b}$
\end{myproof} \noindent
Let $\pi$ be an arbitrary simple path and let node $n$ be one of its on-path nodes. We want to show that increasing the \pairalloc{} \T{n}{i}{j} by some amount $\delta > 0$ does not decrease the allocation calculated by \gma{} for path $\pi$.
Let $\gmaG{\pi}$ be the formula from \cref{eq:gma} and $\xold$ be the argument of its minimum before increasing \T{n}{i}{j}, and let $\gmaGhat{\pi}$ be the formula from \cref{eq:gma} and $\xnew$ the argument of its minimum after increasing~\T{n}{i}{j}. We can distinguish three cases and write $\gmaGhat{\pi}$ as follows:
\begin{subequations}
  \begin{align}
  \xnew< n: \nonumber\\ \gmaGhat{\pi}
  &=\prod_{k=1}^{\xnew -1} \frac{\T{k}{i}{j}}{\conv{k}{j}}\cdot\T{\xnew}{i}{j}\cdot\prod_{k=\xnew +1}^{n-1} \frac{\T{k}{i}{j}}{\divr{k}{i}} \cdot\frac{\T{n}{i}{j}+\delta}{\divr{n}{i}+\delta}\cdot\prod_{k=n +1}^{\ell} \frac{\T{k}{i}{j}}{\divr{k}{i}}
  \label{eq:monotonicity-x-smaller-n}\\
  \xnew = n: \nonumber\\ \gmaGhat{\pi}
  &=\prod_{k=1}^{\xnew -1} \frac{\T{k}{i}{j}}{\conv{k}{j}}\cdot(\T{n}{i}{j}+\delta)\cdot\prod_{k=\xnew +1}^{\ell} \frac{\T{k}{i}{j}}{\divr{k}{i}}
  \label{eq:monotonicity-x-equal-n} \\
  \xnew > n: \nonumber\\ \gmaGhat{\pi}
  &=\prod_{k=1}^{n-1} \frac{\T{k}{i}{j}}{\conv{k}{j}} \cdot\frac{\T{n}{i}{j}+\delta}{\conv{n}{j}+\delta}\cdot\prod_{k=n+1}^{\xnew -1} \frac{\T{k}{i}{j}}{\conv{k}{j}}\cdot\T{\xnew}{i}{j}\cdot\prod_{k=\xnew +1}^{\ell} \frac{\T{k}{i}{j}}{\divr{k}{i}}
  \label{eq:monotonicity-x-greater-n}
  \end{align}
\end{subequations}
The following derivation holds for all of the cases above and directly proves monotonicity:
\begin{subequations}
  \begin{align}
  \gmaGhat{\pi}
  &\geq\prod_{k=1}^{\xnew -1} \frac{\T{k}{i}{j}}{\conv{k}{j}}\cdot\T{\xnew}{i}{j}\cdot\prod_{k=\xnew +1}^{\ell} \frac{\T{k}{i}{j}}{\divr{k}{i}}
  \label{eq:monotonicity-using-lemma}\\
  &\geq\prod_{k=1}^{\xold -1} \frac{\T{k}{i}{j}}{\conv{k}{j}}\cdot\T{\xold}{i}{j}\cdot\prod_{k=\xold +1}^{\ell} \frac{\T{k}{i}{j}}{\divr{k}{i}}
  = \gmaG{\pi}
  \label{eq:monotonicity-using-argmin}
  \end{align}
\end{subequations}
To get \cref{eq:monotonicity-using-lemma}, we applied \cref{le:increase-by-delta} to \cref{eq:monotonicity-x-smaller-n,eq:monotonicity-x-greater-n} and the assumption that $\delta > 0$ to \cref{eq:monotonicity-x-equal-n}.
In the step from \cref{eq:monotonicity-using-lemma} to \cref{eq:monotonicity-using-argmin}, we used the fact that $\xold$ is the argument of the minimum of \cref{eq:gma}.
\qed

\section{Extensibility}
\label{app:extensibility}
\label{app:F}

In real-world implementations of resource-allocation protocols, messages need to be sent on the desired paths in order to discover information about the \allocmatrix{s} of the on-path nodes. To avoid unnecessary communication overhead, we want intermediate nodes to be able to drop allocation messages if the preliminary allocation up to such a node is below a certain threshold. This is captured by the following supplementary property:

\begin{enumerate}[goalsS4, align=left]
  \item \textbf{Extensibility:} Algorithm $\pcap{}$ should allow to calculate a preliminary allocation for every preliminary prefix-path~$\pi^z$ of length~$z$ of some terminated path~$\pi$ ($\pi^z = [ (i^1, j^1), (i^2, j^2),\dots, (i^{z}, j^{z}) ]$ for $1 \leq z < \ell$), where we require that $\pcapbrackets{\pi^1} \geq \pcapbrackets{\pi^2} \geq \dots \geq \pcapbrackets{\pi}$.
    \label{req-extensibility}
\end{enumerate}

\begin{thm}
\gma{} satisfies property~\labelcref{req-extensibility}.
\end{thm}
\begin{myproof}
For every prefix-path $\pi^z = [ (i^1, j^1), (i^2, j^2),\dots, (i^{z}, j^{z}) ]$ ($2 \leq z \leq \ell$) of some terminated path $\pi$, we have
\begin{subequations}
  \begin{align}
  \gmaG{\pi^z} ~
  &=~ \left(\prod_{k=1}^{z} \T{k}{i}{j}\right)\cdot \min_{1\leq x\leq z} \left(\prod_{k=1}^{x-1} \frac{1}{\conv{k}{j}}\cdot \prod_{k=x+1}^{z} \frac{1}{\divr{k}{i}} \right)
  \label{eq:extensibility-start}\\
  &=~ \left(\prod_{k=1}^{z} \T{k}{i}{j}\right)\cdot\min\biggl(\min_{1\leq x\leq z-1} \left(\prod_{k=1}^{x-1} \frac{1}{\conv{k}{j}}\cdot \prod_{k=x+1}^{z-1} \frac{1}{\divr{k}{i}} \right)\nonumber \\
  &\hphantom{{}={}}\cdot\frac{1}{\divr{z}{i}},~\prod_{k=1}^{z-1}\frac{1}{\conv{k}{j}}\biggr)
  \label{eq:extensibility-extended-min}\\
  &\leq \left(\prod_{k=1}^{z} \T{k}{i}{j}\right)\cdot \min_{1\leq x\leq z-1} \left(\prod_{k=1}^{x-1} \frac{1}{\conv{k}{j}}\cdot \prod_{k=x+1}^{z-1} \frac{1}{\divr{k}{i}} \right)\cdot\frac{1}{\divr{z}{i}}\\
  &=\frac{\T{z}{i}{j}}{\divr{z}{i}}\cdot\gmaG{\pi^{z-1}}\\
  &\leq~ \gmaG{\pi^{z-1}}.
  \end{align}
\end{subequations}
We started with \cref{eq:gma} and in the step from \cref{eq:extensibility-start} to \cref{eq:extensibility-extended-min} used the fact that $\displaystyle\min_{1\leq x\leq z}\left(f(x)\right) = \min\left(\min_{1\leq x\leq z-1}\left(f(x)\right), f(z)\right)$.
The last inequality follows from \cref{eq:div_con}.
\end{myproof}

\section{Simulation details}
\label{app:simulation_details}
\label{app:G}

\begin{figure}[!htpb]
  \centering
        \includegraphics[width=0.6\linewidth]{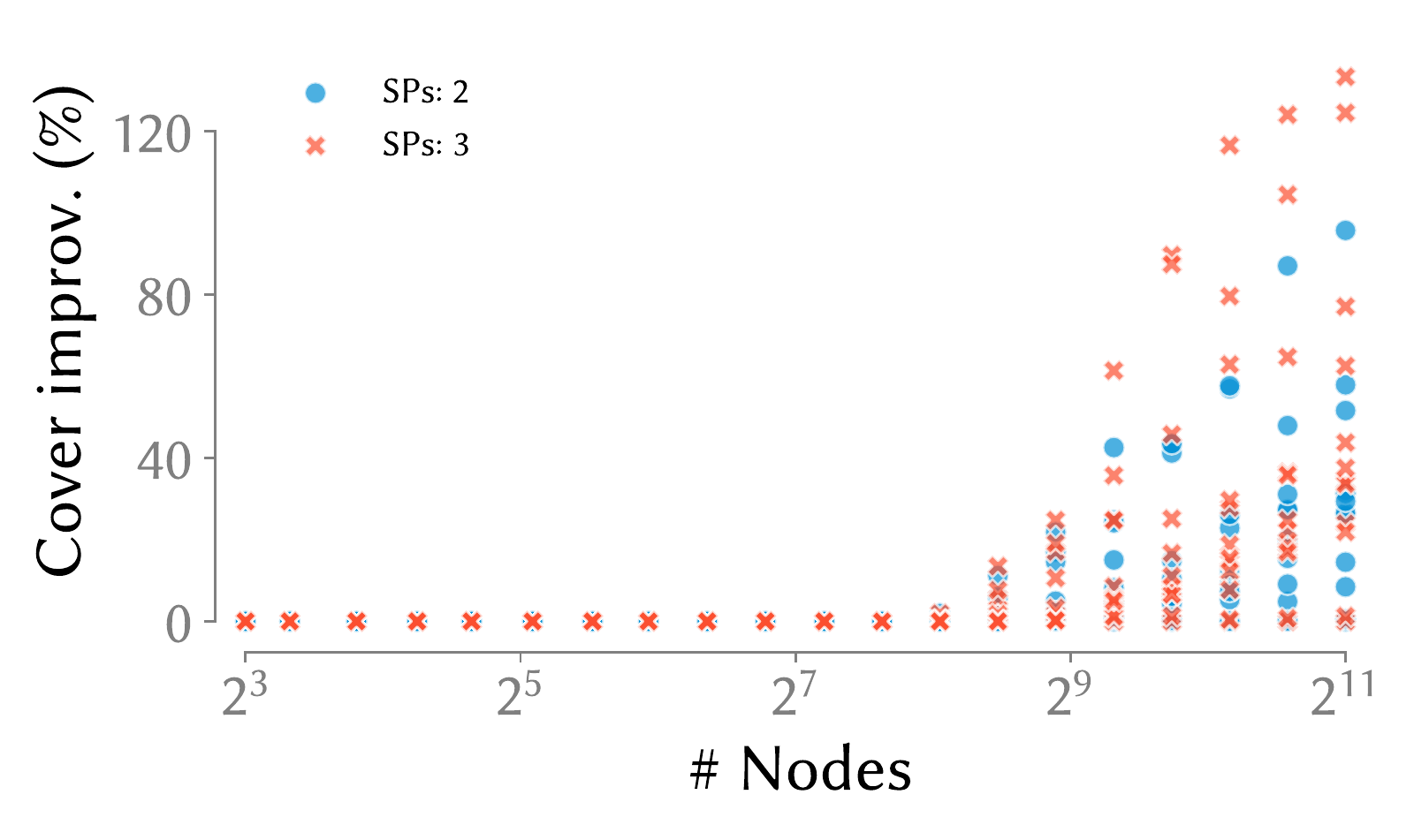}
        \caption{\textbf{Improvement in the median \protect\acover{$\boldsymbol{10^{-4}}$}}
         when using the 2- and 3-shortest path selection schemes instead of the single-shortest selection scheme.}
         \label{fig:barabasi_improvement}
         \label{fig:3}
 \end{figure}

\begin{figure}[!htpb]
  \centering
  \includegraphics[height=6.5cm]{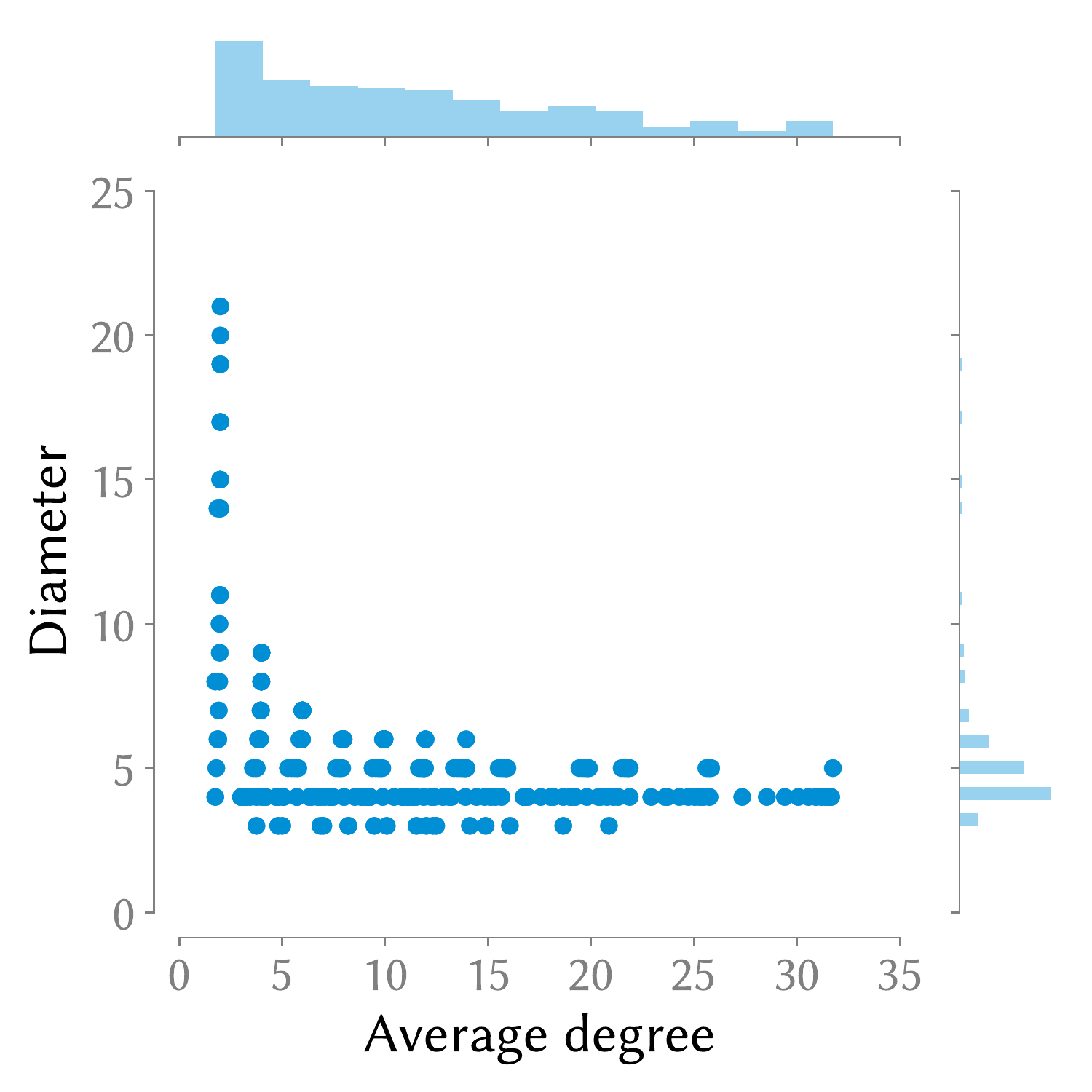}
  \caption{\textbf{Simulated graphs by degree and diameter.} As the marginals show, graphs span a wide range of values in diameter and average node degree.}
  \label{fig:degree_diameter}
  \label{fig:4}
\end{figure}

In the Barabási--Albert model, average degree and diameter are controlled by a \emph{preferential attachment} parameter, and the total number of nodes. A higher preferential attachment will yield graphs with higher average degree and smaller diameter. We vary these two parameters to obtain $275$ random graphs, with the number of nodes varying exponentially from \num{8} to $2048$, and the attachment from \num{1} to \num{32} (the attachment always has to be smaller than the number of nodes).

The relation between the average degree and the diameter of the resulting topologies is visualized in~\cref{fig:degree_diameter}.
\Cref{fig:barabasi_improvement,fig:barabasi_cover_diameter} show additional evaluation results. 
\Cref{fig:single_graph} shows the detail of the \protect\acover{$10^{-4}$} for each node in the graph highlighted in \cref{fig:barabasi_cover_nodes,fig:barabasi_cover_diameter}.

\begin{figure}[!htpb]
        \centering
  \includegraphics[width=0.6\linewidth]{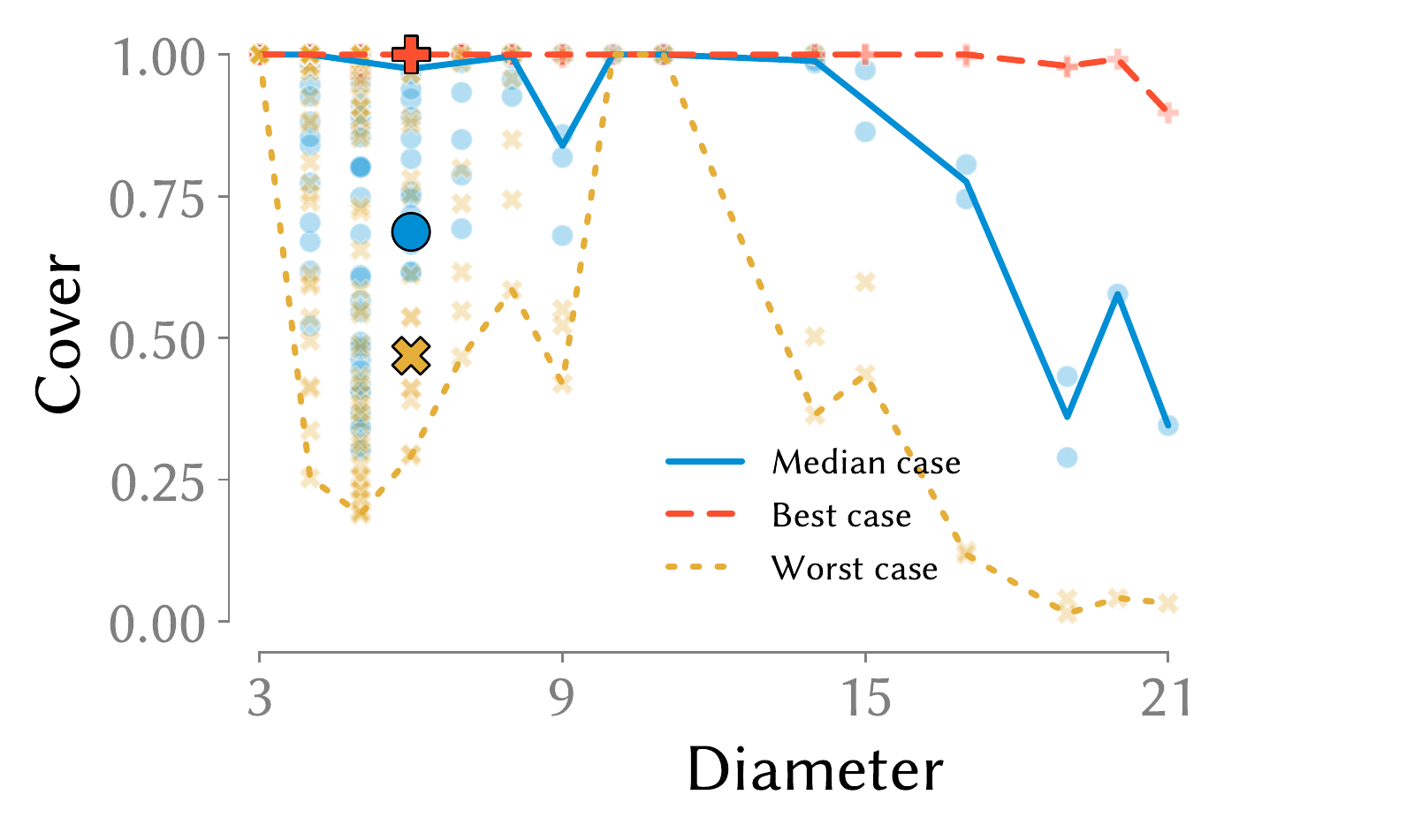}
    \caption{\textbf{Minimum, maximum, and median single-path \protect\acover{$\boldsymbol{10^{-4}}$} breakdown.}
    The highlighted markers show the maximum \ding{58}, median \ding{108}, and minimum \ding{54} cover for one specific graph.}
    \label{fig:barabasi_cover_diameter}
    \label{fig:5}
\end{figure}

\begin{figure}[!htpb]
        \centering
        \includegraphics[width=0.6\linewidth]{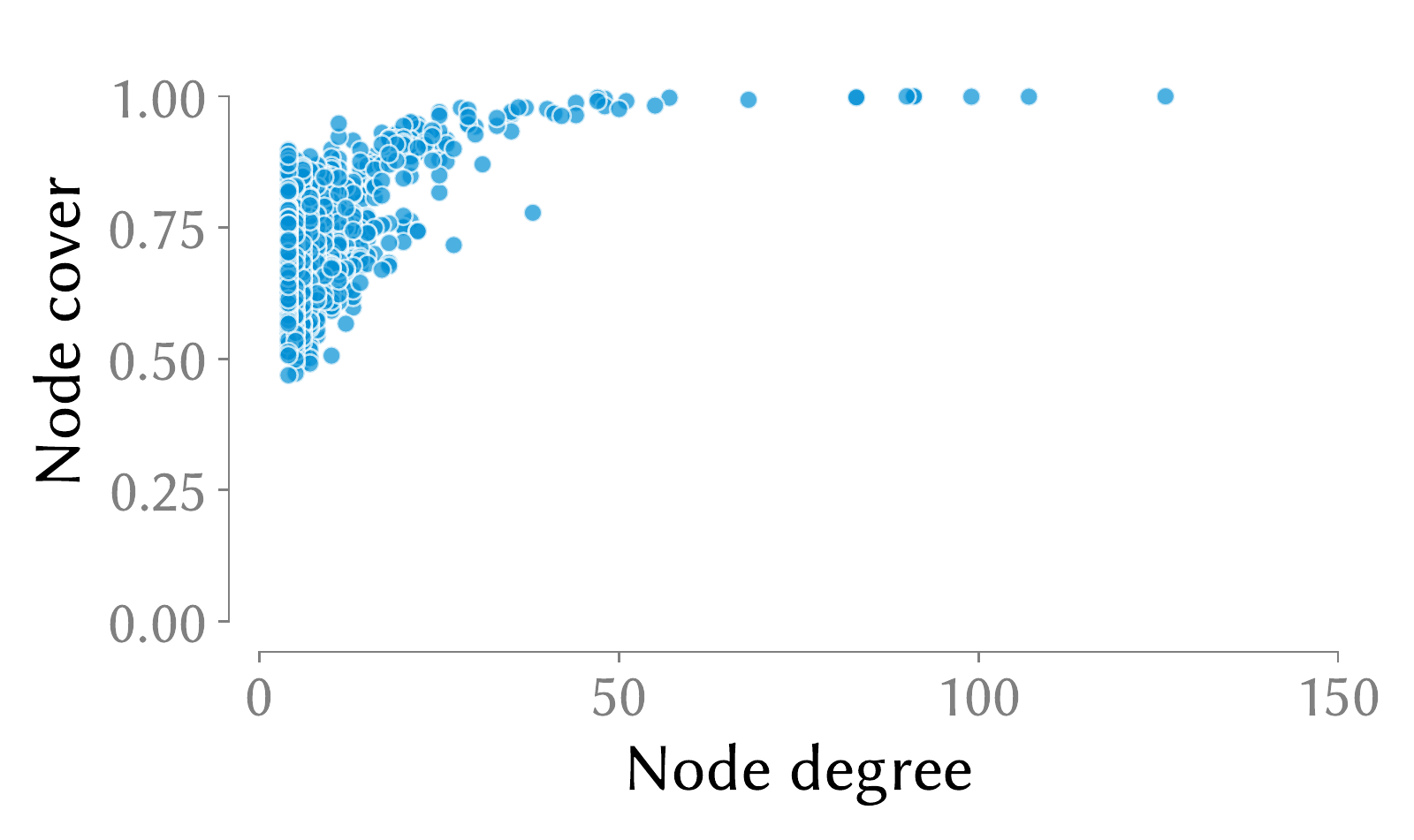}
        \caption{\textbf{Cover and degree of a single graph.}
            Each point is a node of the graph highlighted in \cref{fig:barabasi_cover_nodes,fig:barabasi_cover_diameter}.
        }
        \label{fig:single_graph}
        \label{fig:6}
\end{figure}

\FloatBarrier

\end{document}